\documentstyle[aps,preprint]{revtex}
\begin{document}

\parbox{0.8\textwidth}{{\Large\bf Perturbation theory for nearly 
integrable multi-component  nonlinear PDEs}}\\
\begin{center}
\parbox{0.8\textwidth}{{\sf  V. S. Shchesnovich}${}^{a)}$\\
{\it Department of Mathematics and Applied Mathematics, University of 
Cape Town, Private Bag 7701 Rondebosch, South Africa}${}^{b)}$\\[-0.5cm]

(Received \ \ \ \ \ \ \ \ \ \ \ \ \ \ \ \ \ \ \ \ \ \ \  \ \ )\\[-0.5cm]

The Riemann-Hilbert problem associated with 
the  integrable PDE is used as a nonlinear transformation of the 
nearly integrable PDE to the spectral space. The temporal evolution 
of the spectral  data is derived with account for arbitrary perturbations 
and is given in the form of 
exact equations, which generate the sequence of  approximate ODEs 
in successive orders with respect to the perturbation. For vector nearly 
integrable PDEs, embracing the vector NLS and complex modified KdV 
equations, the main result is formulated  in a theorem. For a 
single vector soliton the evolution equations for the soliton 
parameters and first-order radiation are given in explicit form.}
\end{center}
\vfill
\hrule
\medskip

\noindent
{{\small\rm  ${}^{a)}$Electronic mail: 
valery@maths.uct.ac.za}
\medskip

\noindent
{\small\rm ${}^{b)}$On leave from the Division for 
Optical Problems in Information Technologies, National Academy of 
Sciences of Belarus, Zhodinskaya St. 1/2, 220141 Minsk, Belarus} }
\newpage

\section{Introduction}

Multi-component (or coupled)  nonlinear PDEs had been a subject of 
considerable interest for many years (see, for instance, 
Ref.~1, and references therein). Recent revival of interest in 
multi-component PDEs is due to new discoveries and technological advances in 
nonlinear optics and physics of condensed matter. 
An important example is the incoherent spatial optical  solitons, 
or self-trapped spatially incoherent light beams, recently 
experimentally observed in nonlinear media [2], which are 
described by the multi-component  nonlinear Schr\"odinger (NLS) 
equations [3].  Another example of possible application of the 
coupled NLS equations is the creation and dynamics of solitary waves in the 
multispecies Bose-Einstein condensates [4]. Similar models of 
coupled nonlinear PDEs appear in the wavelength division multiplexing, i.e., 
copropagation of pulses in an optical fibre on beams with different 
wavelengthes [5-7] and in other important applications [8,9]. 

Some of the multi-component models are integrable. Integrable 
multi-component PDEs have another specific feature, which makes them 
important for applications as zero-approximation models for analytical 
description of the real phenomena. It has been known for quite some 
time that dimensional  reductions of matrix generalizations of the 
integrable PDEs, such as the NLS and KdV equations, can produce a variety of 
new integrable equations [10]. For instance, some of the coupled 
NLS equations are integrable reductions of the general matrix NLS equation.  
The $N$-dimensional matrix NLS equation is the simplest integrable PDE 
associated with the  $(N+1)$-dimensional
Zakharov-Shabat spectral problem [11-13].  
Recently  a variety of integrable  coupled higher-order NLS equations 
was discovered [14-16], which are important in view of 
applications to the soliton propagation of sub-picosecond pulses in optical 
fibre [17-19]. Some of these integrable PDEs arise as 
dimensional reductions of the matrix complex modified KdV (cmKdV) equation, 
which is also associated with the Zakharov-Shabat spectral problem.

In most cases, the multi-component PDE is not 
integrable.  However, frequently the terms 
destroying integrability contain small parameters  and  
the non-integrable equation can be considered  as a perturbation of the 
integrable one. In this case, a perturbation theory is required for  
analytical description of the effect of small perturbations.  For instance, 
one is especially  interested in the dynamics of solitons in nearly  
integrable PDEs. Soliton solutions to multi-component equations have 
many parameters, and their evolution may exhibit a variety of new 
interesting regimes. Therefore, it is necessary to have at hand a 
perturbation theory for  multi-component nearly integrable equations. 
Such perturbation theory is developed in the present paper. 

Perturbation theory  for nearly integrable PDEs  has a long 
history [20-45]. There are two basic approaches in the 
perturbation theory based on the IST method.  The first one 
originated in works of Kaup [20] and Karpman and Maslov [21],
where the perturbation theory was developed for nearly integrable
PDEs associated in the integrable limit with the $2\times 2$ matrix
Zakharov-Shabat spectral problem. Quite different approach originated from 
works [22-26]. It was found that an integro-differential 
operator, generating the whole hierarchy of integrable PDEs related to a 
given spectral problem, called the recursion operator,  has a complete set 
of eigenfunctions, which can be used for the perturbation expansion.  
Several other methods, not related to the IST, were applied 
for description of the perturbed soliton dynamics. For instance, a method 
based on the Green functions was developed in Ref.~29. The IST-independent 
perturbation theories for solitons are usually referred to as the direct 
perturbation theories  (see for instance, Ref.~40 and references therein). 
However, notwithstanding the long history of the perturbation theory, with 
rare exceptions, only the $2\times 2$ matrix spectral problems were 
considered.  It was noted that construction of the perturbation theory for 
higher-dimensional matrix spectral problems along the standard approach 
becomes technically more involved.

To overcome technical difficulties of the standard approach when dealing 
with multi-component PDEs, the method based on the 
Riemann-Hilbert (RH) problem was proposed in Ref.~41, where the 
perturbation theory was developed for the  Zakharov-Shabat 
spectral problem of an arbitrary matrix dimension. The RH problem 
was used before for construction of the perturbation theory for the 
Landau-Lifschitz equation [30], the NLS and Maxwell-Bloch 
equations [34], which are integrable by the $2\times2$ matrix spectral 
problem. The approach of Ref.~41 was applied to the Manakov system [42] 
(i.e., the two-component NLS equation), modified NLS equation [43] 
and massive Thirring model [44].  These examples demonstrate that the 
perturbation theory based on the RH problem always works. Recently, the RH 
problem was applied to nearly integrable equations on the half line, arising 
from the singular dispersion relations [45]. In Refs.~41-45 the 
perturbation-induced evolution equations for the spectral data were derived 
with the help of  some matrix functional (below, the evolution functional
$\Pi(x,t,k)$). 
It is important to emphasize that 
the form of the evolution functional is invariant  under the gauge 
transformations of the considered PDE [43]. Thus, once constructed, 
the evolution functional is valid not only for the whole  hierarchy of PDEs 
associated with a given spectral problem, but also  for their images under 
the gauge transformations. Writing the dispersion law, generating the 
spectral problem, in an abstract form $\Lambda(k)$ (see the next section for 
details) we discover that the form of the evolution functional remains 
invariant under the change of the {\it spectral problem} as well. This 
invariance trivially extends to the  general initial-boundary value 
problems. For instance, for the half-line, where one would expect a 
difference, we have found that the evolution functional has similar 
form [45]. Therefore, it seems that the approach based on the 
evolution functional is universal for construction of the perturbation 
theories for nearly integrable PDEs. It is also technically simple.  
Derivation of the perturbation-induced evolution equations for the spectral 
data using the evolution functional reduces to calculation of integrals.

This paper is a further development of Ref.~41. The previous 
results are substantially advanced. In particular, the evolution 
equations for the spectral data  are considerably simplified with the help 
of some identities found for the evolution functional.   We start section 
\ref{secLax} with a brief discussion of the multi-component integrable PDEs 
associated with the  Zakharov-Shabat spectral 
problem. We consider two examples, the  matrix NLS and cmKdV 
equations, however, our approach is valid for many other multi-component 
PDEs. We have not made an attempt  to give  a complete exposition of   
the properties of integrable equations. We need only the Lax 
representation. Hence a way of deriving the Lax pair for an 
 integrable PDE from the  dispersion relation of its linearization is 
briefly indicated. For completeness of the exposition, a detailed 
derivation of the RH problem is given in section \ref{secRH}. Solution of 
the RH problem for multi-component equations involves some technicalities, 
which are discussed  and detailed derivations are provided in the 
Appendices.  We derive evolution equations for the 
spectral data with account for perturbations in section \ref{pert}.  There, 
for an important special case of the vector nearly integrable  PDEs 
the main result of this paper is formulated in a theorem. In the 
final section \ref{exmpl} the equations of the first-order perturbation 
theory for a single vector soliton are given in explicit form.
 
\section{Preliminaries: Integrable multi-component nonlinear PDEs}
\label{secLax}

Here we briefly discuss integrable PDEs with emphasis on the multi-component 
equations whose reductions  are important for applications. 
In particular, we consider the  matrix  
nonlinear Schr\"odinger and complex modified  Korteweg-de Vries equations.
We do not try to review this  subject, for general considerations the reader 
can consult, for  instance, Refs.~1, 10-13, 46-53 and the references 
therein.  The purpose of this section is to remind some of the basic notions 
in the  IST method.  Though the approach below can be applied to any 
nonlinear PDE solvable by the RH problem, we restrict the consideration to 
the integrable equations associated with the $N$-dimensional Zakharov-Shabat 
spectral problem (\ref{lax1}). 

Consider the integrable PDEs which arise as the compatibility condition   
for the following $N\times N$ matrix linear system (Lax pair)
\begin{equation}
\partial_x \Phi = ik[A,\Phi] + iQ(x,t)\Phi\equiv \Phi\Lambda(k)+U(x,t,k)\Phi,
\label{lax1}\end{equation}
\begin{equation}
\partial_t \Phi = {i\omega(k)} \Phi A + V(x,t,k)\Phi\equiv\Phi\Omega(k)+V(x,t,k)\Phi,
\label{lax2}\end{equation}
with 
\[
A=\left(\begin{array}{cc} I_n & 0\\ 0 & -I_{N-n} \end{array}\right),\quad 
Q=\left(\begin{array}{cc} 0& \bf{q}\\ \overline{\bf{q}} & 0 \end{array}\right),
\]
\[
 \bf{q} = \left(\begin{array}{cccc} q_{11}& q_{12}&\ldots& q_{1,{N-n}} \\ 
 q_{21}& q_{22}&\ldots& q_{2,{N-n}} \\
\vdots & \vdots &  & \vdots \\
q_{n1}& q_{n2}&\ldots& q_{n,{N-n}}  \end{array}\right),\quad 
\overline{\bf{q}} =  \left(\begin{array}{cccc} \overline{q}_{11}& 
\overline{q}_{21}&\ldots& \overline{q}_{n1} \\ 
 \overline{q}_{12}& \overline{q}_{22}&\ldots& \overline{q}_{n2} \\
\vdots & \vdots &  & \vdots \\
\overline{q}_{1,{N-n}}& \overline{q}_{2n}&\ldots& \overline{q}_{n,{N-n}}  
\end{array}\right),
\]
where $\Lambda(k) = - ikA$ and $\Omega = i\omega(k)A$ are the  
dispersion laws, $Q$ is called the potential. Here the overline 
does not mean complex conjugation by default, e.g., in general, the 
functions $q_{ij}$ and $\overline{q}_{ij}$ are not considered as complex 
conjugate to each other. When the overline does denote complex 
conjugation in the text below, each time it will be specially indicated. 
This special case corresponds to the Hermitian potential $Q$, $Q^{\dag}=Q$ 
or ${\bf q}^\dagger = {\overline{\bf q}}$, and it will be referred to as the 
involution. The temporal evolution equation (\ref{lax2}) is specified by
choice of the dispersion relation $\omega(k)$ in the following manner. 
For simplicity, let 
the dispersion relation be polynomial  
$\omega(k)=\sum_{p=1}^{M}w_p k^p$, then \begin{equation}
V(k) = -{\cal P}\{\Phi\Omega\Phi^{-1}\}\equiv -\Omega(k)
+\sum_{p=0}^{M-1}V_pk^p.
\label{V}\end{equation}
Here the matrix function $\Phi(k)$ is expanded into the asymptotic 
series: \[
\Phi(k)=I+k^{-1}\Phi^{(1)}+k^{-2}\Phi^{(2)}+..., \quad k\to \infty,
\]
and the operator ${\cal P}$ takes the polynomial in $k$  part of 
$\Phi\Omega\Phi^{-1}$ on the asymptotics.  For example,
the Zakharov-Shabat spectral problem (\ref{lax1}) is 
derived in this way
\[
U=-{\cal P}\{\Phi\Lambda\Phi^{-1}\}=ikA+i[\Phi^{(1)},A],
\]
with the obvious relation 
\begin{equation}
Q=[\Phi^{(1)},A].
\label{Q}\end{equation}
Hence, the Lax pair satisfies the property 
$\text{Tr}U = -\text{Tr}\Lambda$ and  $\text{Tr}V = - \text{Tr}\Omega$.

The integrable nonlinear PDE related  to the Lax pair 
(\ref{lax1})-(\ref{lax2}) is given by the compatibility condition (in our 
case, polynomial in $k$) 
\begin{equation}
i\partial_t Q  -\partial_x V + [ikA+iQ,V] = 0
\label{NLEE}\end{equation}
via setting $k=0$, while the positive powers of $k$ supply the 
expressions  of  the coefficients $V_p$ in (\ref{V}) as functions of the 
potential $Q$ and its  $x$-derivatives, $V_p=V_p(Q,Q_x,Q_{xx},...)$. For 
instance, choosing $\omega(k)=2k^2$ we obtain the well-known matrix 
nonlinear Schr\"odinger equation. Indeed, in this case \[
V=-2ik^2A-2ikQ-AQ_x+iAQ^2
\]
and the equation (\ref{NLEE}) becomes 
\begin{equation}
iAQ_t +  Q_{xx}+2Q^3 = 0.
\label{NLS}\end{equation}
For a complete classification of the matrix integrable NLS equations with various reductions 
to Hermitian symmetric spaces consult Ref.~10. A particular 
important case of equation (\ref{NLS}) is the vector NLS equation, a 
generalization of the  two-component vector NLS, which was shown to be 
integrable by Manakov [12]. The vector NLS equation 
corresponds to the Hermitian potential and the reduction $n=N-1$ (see the 
expression for $A$). In this case we have \begin{equation}
Q = \left(\begin{array}{cccc} 0 & \ldots & 0 & q_1 \\ 
\vdots &   & \vdots & \vdots \\ 0 & \ldots & 0 & q_{n} \\ 
\overline{q}_{1}& \ldots & \overline{q}_{n} & 0  \end{array}\right)
\label{specialQ}\end{equation}
and matrix equation (\ref{NLS}) becomes the vector NLS equation:
\begin{equation}
i\partial_t q_l +\partial_{x}^2 q_l +2\left(\sum_{j=1}^{n}|q_j|^2\right)q_l=0,\quad l=1,...,n.
\label{vectorNLS}\end{equation}

Let us consider another important example of multi-component nonlinear 
integrable equations. It is given by setting $\omega(k)= 4k^3$. After 
simple computation we get \[
V = -4ik^3A+4ik^2Q+2ikA\left(iQ_x+Q^2\right)-iQ_{xx}+[Q,Q_x]-2iQ^3,
\]
which produces the matrix cmKdV equation
\begin{equation}
Q_t + Q_{xxx} +3\left(Q^2Q_x + Q_xQ^2\right) = 0.
\label{CMKdV}\end{equation}

Combining together the considered two dispersion laws, i.e., 
letting $\omega(k) = 4i\epsilon k^3 +2i\beta k^2$, 
one can obtain 
\[
iQ_t + \beta A \left( Q_{xx} + 2Q^3\right) + i\epsilon Q_{xxx} + 
3i\epsilon \left( Q^2Q_x + Q_xQ^2\right)
= 0,
\]
a special case of the (generally, non-integrable) matrix higher-order NLS  
equation
 \begin{equation}
iE_z + A(\alpha_1 E_{\tau\tau} + \alpha_2 E^3) + i \left\{ \alpha_3 
E_{\tau\tau\tau} 
 + \alpha_4 (E^3)_\tau +\alpha_5 (E^2)_\tau E \right\} = 0,
\label{HONLS}\end{equation}

Our examples are just illustrative. There is many other integrable matrix 
PDEs which we do not mention here. However, the perturbation theory 
developed in section \ref{pert}  applies to such PDEs also. 

\section{Riemann-Hilbert problem for multi-component PDEs}
\label{secRH}

In this section we derive the RH 
problem for the multi-component integrable PDEs and discuss its solution 
(for more details consult Refs.~46, 54-57).  We are interested in the 
initial-value problem for nonlinear PDEs on the whole real line with the 
asymptotically vanishing conditions (Cauchy problem) $q_{ij}\to 0$ as 
$|x|\to \infty$. The vanishing asymptotics allows us to concentrate entirely 
on the spectral equation (\ref{lax1}) in derivation of the RH problem, 
while $t$-dependence enters parametrically in our approach (for the 
RH problem for initial-boundary value problems consult Ref.~47). Below we 
omit the explicit $t$-dependence for simplicity of the presentation. To 
begin with, let us summarize the properties of the spectral problem 
(\ref{lax1}). Define the following $N\times N$ matrix projectors 
\begin{equation}
H_1 = \text{diag}(I_n,0),\quad H_2 =  \text{diag}(0,I_{N-n}).
\label{Hs}\end{equation}
Then $A = H_1 -H_2$ and any matrix  can be decomposed into the sum 
of two matrices, commuting and anti-commuting with $A$:
\begin{equation}
\Phi = \Phi^{(c)} +  \Phi^{(a)},\quad  \Phi^{(c)}  = H_1 \Phi H_1 + H_2 \Phi H_2,\quad
\Phi^{(a)}  = H_1 \Phi H_2 + H_2 \Phi H_1,
\label{}\end{equation}
where $[A,\Phi^{(c)}] = 0$ and $\{A, \Phi^{(a)}\} = 0$. 
We will use the block-index  notations 
for the decomposition of matrix $\Phi$ with respect to the projectors
$H_1$ and $H_2$:
\[
\Phi = \left(\begin{array}{cc} \Phi_{I,I} &  \Phi_{I,II}\\
\Phi_{II,I} & \Phi_{II,II}\end{array}\right).
\]

The RH  problem is a one-to-one mapping 
(nonlinear Fourier transform) between the set of smooth (e.g.,  belonging to 
the Schwartz space) potentials $Q(x)$ and some set of the spectral data. 
To identify the RH problem one must construct two solutions, one,  
$\Phi_+(x,k)$, to the spectral equation (\ref{lax1}) and the other,
$\Phi_-^{-1}(x,k)$, to the adjoint equation 
(the second function is inverse of some matrix function satisfying 
(\ref{lax1}), hence the ``-1" in its definition), holomorphic with 
respect to the spectral parameter $k$ in some complementary domains covering 
the whole complex $k$-plane.   Such solutions can be built from 
the columns and rows of the Jost solutions $J_\pm$, i.e., solutions 
defined by the asymptotic conditions: $J_\pm(x,k)\to I$ as $x\to\pm\infty$.

As $\text{Tr} Q = 0$, letting $x\to \pm\infty$ we conclude that 
det$J_\pm = 1$. Hence, the columns of either of the two Jost solutions  
give a linear basis in the space of solutions of the spectral equation. The 
inverse matrices $J_\pm^{-1}$ satisfy the adjoint spectral equation:
\begin{equation}
\partial_x \widetilde\Phi= ik [A, \widetilde\Phi] - i\widetilde\Phi Q.
\label{adj_lax1}\end{equation}
Each column of $J_\pm(x,k)$ and, respectively,  row of $J_\pm^{-1}(x,k)$ 
is holomorphic and bounded in either upper (Im$k\ge0$) or lower 
(Im$k\le0$) half of the complex $k$-plane. Indeed, this can be easily seen 
from the Volterra integral equations for the Jost matrices written for the 
two blocks of columns: \[
J_\pm(x,k)H_1 = H_1 + i\int\limits_{\pm\infty}^x\text{d}
\xi\, e^{-2ik(x-\xi)H_2} Q(\xi)J_\pm(\xi,k)H_1,
\]
\[
J_\pm(x,k)H_2 = H_2 + i\int\limits_{\pm\infty}^x\text{d}
\xi\, e^{2ik(x-\xi)H_1} Q(\xi)J_\pm(\xi,k)H_2.
\]
The columns of $J_+(k)H_1$ and $ J_-(k) H_2$ are 
holomorphic and bounded in the upper half  of the complex  plane,  while 
columns of  $J_+(k)H_2$ and $J_-(k)H_1$ have the same property in the lower 
half plane.  Similarly, the rows of $J_\pm^{-1}$ satisfy integral equations
\[
H_1J^{-1}_\pm(x,k) = H_1  - i\int\limits_{\pm\infty}^x\text{d}\xi\, 
H_1J^{-1}_\pm(\xi,k) Q(\xi)e^{2ik(x-\xi)H_2},
\]
\[
H_2J^{-1}_\pm(x,k) = H_2  - i\int\limits_{\pm\infty}^x\text{d}\xi\,
H_2J^{-1}_\pm(\xi,k)Q(\xi)  e^{-2ik(x-\xi)H_1},
\]
from which we immediately conclude that $H_1 J^{-1}_+(k)$ and   
$H_2J^{-1}_-(k)$ are holomorphic and bounded  in the upper half plane, while 
$H_2 J^{-1}_+(k)$ and $ H_1J^{-1}_-(k)$ have the same properties in the 
lower half plane.  

On the real line, the Jost solutions are transformed into each 
other by the scattering matrix $S(k)$, 
\begin{equation}
J_-(x,k) = J_+(x,k) e^{ikxA}S(k) e^{-ikxA}.
\label{S}\end{equation}  

For the Hermitian potential, $Q^\dagger = Q$, the matrix  Jost solutions 
satisfy the involution (here the overline means complex conjugation)
\begin{equation}
J_\pm^\dagger(x,k) = J^{-1}_\pm (x,\overline{k}),
\label{invol_J}\end{equation}
where the spectral parameter takes complex values in the upper or lower half 
plane depending on the considered column of  the Jost matrix. In this 
case, the  scattering matrix also satisfies the involution
\begin{equation}
S^\dagger(k) = S^{-1}(k),\quad  k\in \text{Re}.
\label{invol_S}\end{equation}

The holomorphic matrix functions $\Phi_+(k)$ and $\Phi^{-1}_-(k)$, 
satisfying equation (\ref{lax1}) and (\ref{adj_lax1}), respectively, 
are given in terms of columns and rows of the Jost solutions: 
\begin{equation}
\Phi_+ =  J_+H_1 + J_-H_2, \quad   
\Phi^{-1}_-  =   H_1J^{-1}_+  + H_2 J^{-1}_- .
\label{phi+phi-}\end{equation}
The above defined matrix functions are holomorphic and bounded in the upper 
and  lower half planes, respectively. They have the following asymptotics 
\begin{equation}
\Phi_\pm(x,k)\to I,\quad k\to\infty,
\label{phi_asymp_k}\end{equation}
which follow from the Volterra integral equations for the Jost solutions.
For the involution (\ref{invol_J}) the matrices $\Phi_+(k)$  and 
$\Phi^{-1}_-(k)$  are related via
\begin{equation}
\Phi^{\dagger}_+(k) =  \Phi_-^{-1}(\overline{k}). 
\label{invol_phi}\end{equation}
These matrices can be conveniently expressed in terms of  only one Jost 
solution and elements  of the scattering matrix $S(k)$. Indeed,
\[
\Phi_+ = J_+e^{ikxA}(H_1 + SH_2)e^{-ikxA} 
\equiv J_-e^{ikxA}(H_2 + S^{-1}H_1)e^{-ikxA},
\]
\[
\Phi^{-1}_- =  e^{ikxA}(H_1 + H_2S^{-1})e^{-ikxA}J^{-1}_+
\equiv e^{ikxA}(H_2 + H_1S)e^{-ikxA}J^{-1}_-.
\]
Denote $S_+ = H_1 + SH_2$, $S_-  = H_2 + S^{-1}H_1$. These matrices provide  
a factorization of the scattering matrix: $S_+ = SS_-$. Similarly, 
$\overline{S}_+ = H_1 + H_2 S^{-1}$ and  $\overline{S}_- = H_2 + H_1 S$, 
which also factorize the scattering matrix: $\overline{S}_+ S = 
\overline{S}_-$. Then
\begin{mathletters}
\label{rep_factor}
\begin{equation}
\Phi_+ = J_+e^{ikxA}S_+ e^{-ikxA} \equiv J_-e^{ikxA}S_- e^{-ikxA}, 
\label{rep_factor_a}\end{equation}
\begin{equation}
\Phi^{-1}_- =  e^{ikxA}\overline{S}_+e^{-ikxA}J^{-1}_+ 
\equiv e^{ikxA}\overline{S}_-e^{-ikxA}J^{-1}_-.
\label{rep_factor_b}\end{equation}
\end{mathletters}

The factorization matrices have the block-triangular structure. For 
instance, $S_+$ and $\overline{S}_+$ are upper and lower block-triangular, 
respectively: \begin{equation}
S_+ = \left( \begin{array}{cc} I_n  & {\bf b} \\ 0   & {\bf a}  \end{array}\right),\quad
\overline{S}_+ = \left( \begin{array}{cc} I_n  & 0  \\ \overline{{\bf b}}  
& \overline{{\bf a}}    
\end{array}\right),
\label{factors}\end{equation}
where ${\bf b} = S_{I,II}$, ${\bf a} = S_{II,II}$, 
$\overline{{\bf b}} = (S^{-1})_{II,I}$, and $\overline{{\bf a}} = 
(S^{-1})_{II,II}$.  The following identity follows from these definitions
\begin{equation}
\overline{{\bf b}}{\bf b} +\overline{{\bf a}}{\bf a} = I_{N-n}, \quad k\in\text{Re}.
\label{scRH}\end{equation}
For the involution (\ref{invol_phi}) the factorizations  satisfy
\[
S^\dagger_\pm(k) = \overline{S}_\pm(k), \quad k\in \text{Re}.
\]
Hence, $\overline{{\bf b}} = {\bf b}^\dagger$  and $\overline{{\bf a}} = 
{\bf a}^\dagger$ in the case of involution.

Considering the product $\Phi^{-1}_-\Phi_+$ we obtain the  problem of 
analytic factorization of a matrix $G(k)$ given on the real line, i.e., the 
matrix RH problem:
\begin{equation}
\Phi^{-1}_-(x,k)\Phi_+(x,k) = e^{ikxA} G(k) e^{-ikxA}, \quad k\in \text{Re}
\label{RH}\end{equation}
and $\Phi_\pm(x,k)\to I$ for $k\to \infty$. 
Here the matrix $G=\overline{S}_+S_+\equiv\overline{S}_-S_-$ reads
\begin{equation}
 G = \left( \begin{array}{cc} I_n  & {\bf b} \\ \overline{{\bf b}}   & 
I_{N-n}  \end{array}\right).
\label{G}\end{equation}

As it was mentioned, the RH problem is a nonlinear mapping between the 
potential $Q(x)$ and the set of the spectral data, which are necessary for 
unique identification of the solution to (\ref{RH}). For instance, given a 
potential, one can obtain the spectral data by solving the spectral 
equation and its adjoint for $\Phi_+(x,k)$ and $\Phi^{-1}_-(x,k)$. 
Conversely, by asymptotic expansion of $\Phi_\pm(x,k)$ as $k\to\infty$ one 
recovers the potential. The asymptotic expansion of $\Phi_\pm(x,k)$ can be 
derived via integration by parts (in the blocks with $e^{\pm 
2ikx}$) in the Volterra integral equations for $J_\pm$  and 
$J^{-1}_\pm$. It  reads \[
 \Phi_\pm(x,k) =  I - \left( AQ(x) + i\int\limits_{-\infty}^x\text{d}\xi\, 
Q^2(\xi)H_2 +i\int\limits^{\infty}_x\text{d}\xi\, Q^2(\xi)H_1 
\right)\frac{1}{2k} +{\cal O}\left(\frac{1}{k^2}\right). \]
Hence, we obtain (cf. with (\ref{Q}))
\begin{equation}
Q(x) = \lim_{k\to\infty}k[\Phi_+(x,k),A] = \lim_{k\to\infty}k[A,\Phi^{-1}_- 
(x,k)] . \label{QRH}\end{equation}

\subsection*{Solution of the RH problem: Normalization}

Let us discuss the way of solution of  the RH problem. The coordinate 
dependence is not important for this purpose and it is omitted below. In 
general, the determinants  $\det\Phi_+(k)$ and $\det	\Phi^{-1}_-(k)$ have 
zeros and the RH problem is said to be non-regular or with zeros. Note that 
the determinants do  not depend on $x$ as it readily seen from equations 
(\ref{lax1}) and (\ref{adj_lax1}). We consider only the RH problem with 
zero index, i.e., \[
\int\limits_{-\infty}^\infty\text{d}\,\text{ln}\{\det G(k)\} = 0,
\]
assuming that $G(k)$ is non-degenerate. Since $\Phi_\pm\to I$ as $k\to\infty$, 
the index is equal to 
the difference between the number of zeros in the upper and lower half 
planes. Let $k_1,...,k_M$ and $\overline{k}_1, ..., \overline{k}_M$ be zeros 
(where some may be equal) of  $\det\Phi_+(k)$ and $\det	\Phi^{-1}_-(k)$, 
respectively. Two conditions are imposed on the zeros. First, 
the geometric multiplicity of a zero must be equal to its order (which we 
will refer to as the algebraic multiplicity). Here the geometric 
multiplicity of $k_j$ is defined as the dimension $d_j$ of the null space 
of $\Phi_+(k_j)$, i.e., $d_j=N-\text{rank}\Phi_+(k_j)$. Trivially, the two 
multiplicities are equal to 1 in the case  of simple zeros. In 
the case of involution zeros of the RH problem come in complex conjugate 
pairs  (due to formula (\ref{invol_phi})),  with  coinciding algebraic (and 
geometric)  multiplicities within each pair. (Due to the involution the 
index is equal to zero, in this case $\det G(k)$ is real and definite.)  
Second, as we are mainly interested in the Hermitian potentials $Q$, i.e., 
in the case of the involution, we will consider only paired  zeros $k_j$ and 
$\overline{k}_j$, $j=1,...,s\le M$,  whose algebraic multiplicities are 
equal, however without assuming them to be complex conjugate to each other. 
The  algebraic multiplicity $\nu_j$ of $k_j$ always satisfies 
the following inequality (see Appendix~A)
\[
\nu_j\ge N - \text{rank}\Phi_+(k_j)=d_j,
\]
while $\text{rank}\Phi_+ \ge \text{max}(n,N-n)$ by construction  of $\Phi_+$ 
(\ref{phi+phi-}).  For instance, if $n= N-1$, there can be not more than  
one vector in the null space of $\Phi_+(k)$, i.e., $d_j=1$ for all 
$j=1,...,s$. Hence, for this reduction, zeros  of $\det\Phi_+(k)$ must be 
simple to satisfy the equal multiplicity condition.  Similar for zeros of 
$\Phi^{-1}_-(k)$. (More detailed consideration of the multiplicities of 
zeros is placed in Appendix~A.) 

Let $\det \Phi_+(k)$ and $\det \Phi^{-1}_-(k)$ have zeros $k_1,...,k_s$ and 
$\overline{k}_1,...\overline{k}_s$, respectively, with the (algebraic and geometric) 
 multiplicities $\nu_1,..., \nu_s$. 
Exactly $\nu_j$ independent vector-columns $ | p^{(j)}_l \rangle $ 
(vector-rows $ \langle \overline{ p}{}^{(j)}_l |$), where $l=1,...,\nu_j$,
correspond to each zero $k_j$ (respectively, 
$\overline{k}_j$), such that
\begin{equation}
\Phi_+(k_j) | p^{(j)}_l \rangle = 0,\quad  \langle 
\overline{ p}{}^{(j)}_l |\Phi^{-1}_-(\overline{k}_j) = 0. \label{vec_def}
\end{equation} 
To specify a unique solution to the RH problem, additionally to 
the continuous datum $G(k)$, the set of the discrete data $\{ k_j, | 
p^{(j)}_l \rangle, \overline{k}_j, \langle \overline{ p}{}^{(j)}_l |,  
l=1,...,\nu_j, j=1,...,s\}$ must be given.  This becomes evident from the 
fact that the RH problem with zeros (i.e., non-regular) can be regularized, 
i.e., reduced to a modified RH problem without zeros  by factoring 
them out with some rational matrix $\Gamma(k)$:
\begin{equation}
 \Phi_+(k) = \phi_+(k) \Gamma(k),\quad 
 \Phi^{-1}_-(k) = \Gamma^{-1}(k)\phi^{-1}_-(k).
\label{normphi+phi-}\end{equation}
The regularized RH problem reads
\begin{equation}
 \phi^{-1}_-(k)\phi_+(k) =  \Gamma(k) e^{ikxA}G(k)e^{-ikxA}\Gamma^{-1}(k), 
\quad k\in \text{Re}
\label{regRH}\end{equation}
and $\phi_\pm(k)\to I$ as $k\to\infty$. 

The rational matrix functions  
$\Gamma(k)$ and $\Gamma^{-1}(k)$, (for details see Appendix B)
\begin{equation}
\Gamma = I - \sum_{j,l;i,m} \frac{| p^{(i)}_m  
\rangle\left(D^{-1}\right)_{im,jl}\langle \overline{ p}{}^{(j)}_l |}{k - 
\overline{k}_j} ,\quad
\Gamma^{-1} = I + \sum_{j,l;i,m} \frac{
| p^{(j)}_l \rangle\left(D^{-1}\right)_{jl,im} \langle \overline{ 
p}{}^{(i)}_m |}{k - {k}_j}, \label{Gammas} 
\end{equation}
where
\[
D_{im,jl} = \frac{\langle \overline{ p}{}^{(i)}_m | p^{(j)}_l \rangle }{k_j 
- \overline{k}_i},
\]
have the same zeros  as $\Phi_+(k)$ and 
$\Phi^{-1}_-(k)$, respectively, and  the same null spaces:
\[
\Gamma(k_j)| p^{(j)}_l \rangle = 0,\quad  \langle \overline{ p}{}^{(j)}_l |
\Gamma^{-1}(\overline{k}_j) = 0,\quad l=1,...,\nu_j, \quad
j=1,...,s. 
\]
Moreover, $\det \Gamma = 
\prod_{j=1}^s\left(\frac{k-k_j}{k-\overline{k}_j}\right)^{\nu_j}$.
Precisely these properties allow us to factor out zeros of $\Phi_+(x,k)$ and 
$\Phi^{-1}_-(x,k)$. 

It is important to emphasize that there is a freedom in 
choice of the  basis vectors spaning the null spaces. Indeed, it is easy 
to verify that the regularization  matrix $\Gamma$ is invariant under the 
transformations:
\begin{equation}
| p^{(j)}_l \rangle \to \sum_{m=1}^{\nu_j}| p^{(j)}_{m} \rangle M^{(j)}_{m l},
\quad
\langle \overline{ p}{}^{(j)}_l | \to \sum_{m=1}^{\nu_j} \overline{M}{}^{(j)}_{lm}
\langle \overline{ p}{}^{(j)}_{m} |,
\label{invariance}\end{equation}
where $M^{(j)}$  and $\overline{M}{}^{(j)}$ are arbitrary non-degenerate  
$\nu_j\times\nu_j$-matrices. 
For the involution, due to equation (\ref{invol_phi}) we can choose $\langle 
\overline{ p}^{(j)}_l | = | p^{(j)}_l \rangle^\dagger$.
This evidently leads to
\begin{equation}
\Gamma^{\dagger}(k) = \Gamma^{-1}(\overline{k}).
\label{invol_gam}\end{equation}

The $(x,t)$-dependence of the RH data can be found in the following  way. 
The $t$-dependence of the continuous datum (which does not depend on $x$)  
can be derived from equations (\ref{RH}) and (\ref{lax2}).  We obtain 
\begin{equation}
\partial_t G = [G,\Omega].
\label{G_t}\end{equation}
This equation can be cast in a more explicit form:
\begin{equation}
\partial_t {\bf b} = -2i\omega(k) {\bf b},\quad 
\partial_t \overline{{\bf b}} = 2i\omega(k) \overline{{\bf b}}.
\label{b_t}\end{equation}

Recalling that  $\text{Tr}Q=0$ and $\text{Tr}V = -\text{Tr}\Omega$, one can 
verify that the determinants of $\Phi_\pm$ do not depend on the co-ordinates.
Therefore, the zeros are co-ordinate independent as well:
\begin{equation}
\partial_x k_j = \partial_t k_j =0,\quad \partial_x 
\overline{k}_j = \partial_t \overline{k}_j = 0,\quad j=1,...,s.
\label{k_xt}\end{equation}

Now, let us derive the co-ordinate dependence of the vector-parameters. 
Differentiating equation (\ref{vec_def}) we obtain 
\[
\Phi_+(k_j)\left( \partial_x | p^{(j)}_l \rangle  -  ik_j A | p^{(j)}_l \rangle \right) = 0, \quad
\Phi_+(k_j)\left( \partial_t | p^{(j)}_l \rangle  +  \Omega(k_j) | p^{(j)}_l \rangle \right) = 0.
\]
Therefore 
\[
\partial_x | p^{(j)}_l \rangle = ik_jA | p^{(j)}_l \rangle + \sum_{m=1}^{\nu_j}F^{(j)}_{lm}
| p^{(j)}_m\rangle, \quad  
\partial_t| p^{(j)}_l \rangle = -\Omega(k_j) | p^{(j)}_l \rangle + \sum_{m=1}^{\nu_j}K^{(j)}_{lm}
| p^{(j)}_m\rangle,  
\]
where $F$ and $K$ are some matrices.  Using a suitable invariance 
transformation of the type (\ref{invariance}) one can put 
$F=K=0$ without loss of generality. Hence we have 
\begin{equation}
\partial_x | p^{(j)}_l \rangle = ik_jA | p^{(j)}_l \rangle, \quad
\partial_t  | p^{(j)}_l \rangle = -\Omega(k_j) | p^{(j)}_l \rangle. 
\label{p_xt}\end{equation}
Similarly
\begin{equation}
\partial_x  \langle \overline{ p}^{(j)}_l | = - \langle \overline{ p}^{(j)}_l | i\overline{k}_jA,\quad 
\partial_t  \langle \overline{ p}^{(j)}_l | = \langle \overline{ p}^{(j)}_l | \Omega(\overline{k}_j). 
\label{barp_xt}\end{equation}

Some comments are needed on the reconstruction  of the potential $Q$. First 
of all, the soliton part of the potential  is given by the rational matrix 
function $\Gamma(k)$ (\ref{Gammas}). The pure soliton potentials are also 
called reflectionless, they solve the simplest RH problem with $G=I$,
i.e., with zero reflection coefficients: ${\bf b}(k) = \overline{\bf b}(k) =0$. The discrete RH data
have the following meaning. Zeros provide the amplitudes and velocities of 
the solitons, while the null vectors give their initial position, 
polarization and phase parameters. Second, the radiation part of the 
potential is given by the solution to the regularized RH problem 
(\ref{regRH}). The continuous RH data  ${\bf b}(k)$ and $\overline{\bf 
b}(k)$ represent the nonlinear spectral densities of radiation. 
The regular RH problem is equivalent to some matrix 
integral equation of the Fredholm type (for instance, consult 
Ref.~46). Though the solution cannot be given in explicit form, 
its properties can be explored by the standard technique 
of  the theory of Fredholm integral equations.

For the following, it is convenient to introduce 
$x$-independent null vectors $| P^{(j)}_l \rangle$ and 
$\langle \overline{P}^{(j)}_l |  $ setting
\begin{equation}
| p^{(j)}_l \rangle = e^{ik_j x A} | P^{(j)}_l \rangle, \quad
\langle \overline{ p}{}^{(j)}_l | = \langle \overline{P}{}^{(j)}_l | 
e^{-i\overline{k}_j xA}.
\label{def_P}\end{equation}
Then, we have an  $x$-independent set of the RH data, which we will call the
spectral data: 
 $\{ {\bf b}(k), \overline{{\bf b}}(k), 
k_j, | P^{(j)}_l \rangle, \overline{k}_j, \langle \overline{P}{}^{(j)}_l |, 
 l=1,...,\nu_j, j=1,...,s\}$.      
 
As the illustrative  example, consider one-soliton solution. 
It is given by $\Gamma(k)$ having only one pole,
say $k_1=i\eta+\xi$. In this case
\[
Q = (k_1 - \overline{k}_1) 
[A,P_r],\quad P_r = \sum_{l,m=1}^{\nu_1}| p_l \rangle 
\left(\Delta^{-1}\right)_{lm}
\langle \overline{ p}_m |,
\]
with $\Delta_{lm}  = \langle \overline{ p}_l | p_m \rangle$. Here 
 $\langle \overline{ p}_l | \Phi^{-1}_-(\overline{k}_1) = 0$ and 
$\Phi_+(k_1)| p_l \rangle = 0$. In the particular case of $n=N-1$ and 
the Hermitian potential we get the vector soliton 
solution, for which
\[
P_r = \frac{|p\rangle\langle p|}{\langle p|p\rangle},\quad
|p\rangle = \exp\left\{ik_1xA\right\}|P\rangle.
\]
Define complex parameters $C_l= P_l/P_N$. The $t$-dependence of  
$C_l$ then follows from equation (\ref{p_xt}). It can be
accounted for in a convenient way by introducing real $t$-dependent 
parameters $\bar{x}$, and  $\delta_l$, $l=1,...,n$, the soliton position and 
phases of its components. Let $C_l = 
\theta_l e^{2(\eta-i\xi)\bar{x}}$, where
$\theta_l = s_le^{i\delta_l}$.  The amplitudes $s_l$,   
satisfying 
\begin{equation}
\sum_{l=1}^n s_l^2=1,
\label{sdeltas}\end{equation}
describe polarization of the multi-component soliton. 
Integrating equation (\ref{p_xt}) we obtain
\begin{equation}
\bar{x} = \bar{x}_0 + \frac{\text{Im}\{\omega(i\eta 
+\xi)\}t}{\eta},\quad
\delta_l = {\delta_l}_0+\frac{2}{\eta}\text{Im}
\left\{\left(\xi-i\eta\right)\omega(i\eta+\xi)\right\}t. 
\label{tdependence}\end{equation}
The vector soliton  then takes the form 
\begin{equation}
q_l = 2i\eta\theta_l e^{i(\xi/\eta )z }
\text{sech}z
\label{one-sol}\end{equation}
with $z=2\eta(x-\bar{x})$.
For instance, for the vector NLS equation $\omega(k) = 2k^2$ 
and we arrive at the $n$-component generalization of the well-known 
vector soliton solution for the  two-component NLS equation [12].

\section{Perturbation-induced evolution of the spectral data}	
\label{pert}

A perturbation of the integrable PDE, following from the Lax 
representation (\ref{lax1})-(\ref{lax2}), can be written in the form  
\begin{equation}
iQ_t -V_x + [ikA + iQ, V] = R,
\label{perturb_eq}\end{equation}
where $R$  (which is $k$-independent in case of the Zakharov-Shabat 
spectral problem) represents the terms destroying 
 integrablity. The perturbation matrix $R$, similar to the potential $Q$,  
satisfies $[A,R]=0$. For the Hermitian 
potential $Q^\dagger = Q$ we get also $R^\dagger = - R$. 

Below we derive 
evolution equations for the spectral data with account for arbitrary 
perturbation. For simplicity of the presentation we will frequently 
omit the explicit dependence on $x$ and $t$. To distinguish between the 
``integrable" and ``perturbation" contributions to the evolution let us  
assign the variational derivatives to the latter. For instance, the 
perturbation-induced evolution in equation (\ref{perturb_eq}) reads
\[
i\frac{\delta Q}{\delta t} = R,\quad \text{or, explicitly}\quad 
i\frac{\delta {\bf q}}{\delta t} = {\bf r}\quad \text{and}\quad
i\frac{\delta \overline{\bf q}}{\delta t} = -\overline{\bf r},
\]
where ${\bf r} = R_{I,II}$ and $\overline{\bf r} = - R_{II,I}$. For 
Hermitian $Q$,
$ \overline{\bf r}= {\bf r}^\dagger  $.  

We start with derivation of  evolution equations for $\Phi_+$ and 
$\Phi^{-1}_-$, the solution to the RH problem (\ref{RH}). In other terms, we 
are going to derive the generalized Lax pair (see equations (\ref{glax1}) 
and (\ref{ev_phi}) below) for the {\it perturbed} nonlinear PDE 
(\ref{perturb_eq}). Differentiation of  equation (\ref{lax1}) with respect 
to $t$ gives
\[
\partial_x \left(\frac{\delta \Phi}{\delta t}\right) = ik[A,\frac{\delta 
\Phi}{\delta t}] 
+ iQ\frac{\delta \Phi}{\delta t} + R\Phi,
\] 
and consequently
\[
\partial_x \left(\Phi^{-1}\frac{\delta \Phi}{\delta t} \right) = ik[A,\Phi^{-
1}\frac{\delta \Phi}{\delta t}]  
+ \Phi^{-1}R\Phi.
\]
Let us integrate the above formula for $\Phi = J_\pm$. We get
\[
\frac{\delta  J_\pm}{\delta t}(x,k) = J_\pm(x,k)e^{ikxA}\left(\;
\int\limits_{\pm\infty}^x\text{d}\xi\, e^{-ik\xi A}J^{-
1}_\pm(\xi,k)R(\xi)J_\pm(\xi,k)e^{ik\xi A}\right)
e^{-ikx A}.
\]
Here we have used that $\delta J_\pm \to 0$ as $x\to \pm\infty$. Let us 
employ the relation between $J_\pm$ and $\Phi_+$ to rewrite this formula in 
a more convenient form. Using (\ref{rep_factor_a}) we get
\begin{equation}
\frac{\delta J_\pm}{\delta t}(x,k) = J_\pm(x,k)e^{ikxA} S_\pm(k) 
\Upsilon(\pm\infty,x;k) 
S^{-1}_\pm(k) e^{-ikxA}.
\label{varJ1}\end{equation}
In much the same way, using (\ref{rep_factor_b}), we  derive
\begin{equation}
\frac{\delta  J^{-1}_\pm}{\delta t}(x,k)=-e^{ikxA}\overline{S}{}^{-1}_\pm(k)
\overline{\Upsilon} (\pm\infty,x;k)\overline{S}_\pm(k) 
e^{-ikxA}J^{-1}_\pm(x,k).
\label{varJ2}\end{equation}
Here we have introduced the notations:
\begin{equation}
\Upsilon(\pm\infty,x;k) = \int\limits_{\pm\infty}^x\text{d}\xi\, 
e^{-ik\xi A}\Phi^{-1}_+ R \Phi_+ e^{ik\xi A},
\label{Ups}\end{equation}
\begin{equation}
\overline{\Upsilon}(\pm\infty,x;k) = \int\limits_{\pm\infty}^x\text{d}\xi\, e^{-
ik\xi A}
\Phi^{-1}_- R \Phi_- e^{ik\xi A}.
\label{barUps}\end{equation}
These matrix functionals will enter every formula describing 
perturbation-induced evolution of the spectral data. For the involution, 
due to formula (\ref{invol_phi}) these matrix functionals satisfy	
\begin{equation}
\Upsilon^\dagger(\pm\infty,x;k) =  - 
\overline{\Upsilon}(\pm\infty,x,\overline{k}).
\label{invol_ups}\end{equation}
Using  equations  (\ref{varJ1}), (\ref{varJ2}), and the definition of the 
scattering matrix $S$ (\ref{S})  we get 
\[
\frac{\delta  S}{\delta t} = e^{-ikxA}\frac{\delta  }{\delta t}\left( J^{-1}_+ 
J_-\right) e^{ikxA} 
= - \overline{S}{}^{-1}_+\overline{\Upsilon}(\infty,x)\overline{S}_+ S + S S_- 
\Upsilon(- \infty,x)
S^{-1}_- 
\]
\[
= S_+ \Upsilon(- \infty,x) S_-^{-1} + \overline{S}{}^{-1}_+ 
\overline{\Upsilon}(x,\infty)\overline{S}_-.
\]
Setting $x\to\pm\infty$ in this formula produces two simple equivalent 
formulae:
\begin{equation}
\frac{\delta  S(k) }{\delta t} =  S_+(k) \Upsilon(- \infty,\infty;k) S_-^{-1}(k) 
\equiv
 \overline{S}{}^{-1}_+(k) \overline{\Upsilon}(-\infty,\infty;k)\overline{S}_-
(k).
\label{varS}\end{equation}
Now we can easily obtain the  perturbation-induced evolution of the 
solution to the RH problem. Taking into account the definitions  of $S_\pm 
$ and $\overline{S}_\pm$ and (\ref{rep_factor}) we write:
\begin{mathletters}
\label{var_phi}
\[
\frac{\delta  \Phi_+ }{\delta t} 
= \frac{\delta  J_+ }{\delta t}e^{ikxA}S_+e^{-ikxA}  + J_+e^{ikxA} 
\frac{\delta }{\delta t}
( H_1 + SH_2)e^{-ikxA}
\]
\begin{equation}
= \Phi_+e^{ikxA}\left( - \Upsilon (x,\infty) + \Upsilon(-
\infty,\infty)H_2\right)e^{-ikxA}
= \Phi_+e^{ikxA}\Pi e^{-ikxA},
\label{dphi+}\end{equation}
\[
\frac{\delta  \Phi^{-1}_- }{\delta t} 
= e^{ikxA}\overline{S}_+e^{-ikxA}\frac{\delta  J^{-1}_+ }{\delta t}
 + e^{ikxA} \frac{\delta }{\delta t}( H_1 + H_2S^{-1})	J^{-1}_+e^{-ikxA}
\]
\begin{equation}
= e^{ikxA}\left(\overline{\Upsilon}(x,\infty) - H_2\overline{\Upsilon}(-
\infty,\infty)\right)
\Phi^{-1}_-e^{-ikxA} 
=- e^{ikxA}\overline{\Pi}\Phi^{-1}_-e^{-ikxA}. 
\label{dphi-}\end{equation}
\end{mathletters}
The r.h.s.'s of the above formulae contain  the evolution functionals, 
\begin{equation}
\Pi(x,k) = - \Upsilon (x,\infty;k)H_1 + \Upsilon(-\infty,x;k)H_2,
\label{Pi}\end{equation}
\begin{equation}
\overline{\Pi}(x,k) = H_1\overline{\Upsilon}(x,\infty;k) - 
H_2\overline{\Upsilon}(-\infty,x;k),
\label{barPi}\end{equation}
which account for the perturbation-induced evolution of the solution to the 
RH problem. As follows from formula (\ref{invol_ups}), for the case of the 
involution the evolution functionals satisfy
\begin{equation}
\Pi^\dagger(k) = - \overline{\Pi}(\overline{k}).
\label{invol_pi}\end{equation}
From the definitions (\ref{Pi}), (\ref{barPi}) and also (\ref{Ups}), 
(\ref{barUps})  it is easy to 
see that  the matrices $e^{ikxA}\Pi(k) e^{-ikxA}$ and $ 
e^{ikxA}\overline{\Pi}(k)e^{-ikxA}$ are meromorphic and bounded in the 
upper and lower half planes of the $k$-plane, respectively. They have 
simple poles respectively at zeros of $\det\Phi_+(k)$ and  
$\det\Phi^{-1}_-(k)$. 

Let us  write down the generalized Lax representation for the 
perturbed PDE (\ref{perturb_eq}) in terms of $\Phi_+(x,k)$:
\begin{equation}
\partial_x \Phi_+= ik[A,\Phi_+] + iQ\Phi_+, 
\label{glax1}\end{equation}
\begin{equation}
\partial_t \Phi_+ = \Phi_+ \Omega + V\Phi_+ + \Phi_+ e^{ikxA}\Pi e^{-ikxA}.
\label{ev_phi}\end{equation}
It is easy to check by direct calculation that the compatibility condition for 
the above 
linear system is equivalent to the perturbed  equation (\ref{perturb_eq}) 
(indeed,  $\partial_x \Pi = e^{-ikxA}\Phi^{-1}_+R\Phi_+e^{ikxA}$).  

Now, let us derive the perturbation-induced evolution of the spectral data. 
From (\ref{var_phi}) we immediately obtain
\[
\frac{\delta G}{\delta t} = \frac{\delta }{\delta t} \left( e^{-ikxA}\Phi^{-1}_-
\Phi_+e^{ikxA}\right)
= G\Pi - \overline{\Pi}G.
\]
Hence, the complete evolution of the continuous datum reads
\begin{equation}
\partial_t G = [G,\Omega]  + G\Pi - \overline{\Pi}G,
\label{evG}\end{equation}
where we have taken into account the integrable evolution given by (\ref{G_t}). 
It is important to notice that the l.h.s. of (\ref{evG}) does not depend on $x$. 
Therefore, we can put $x\to\pm\infty$ in this equation to simplify it. 
Equation (\ref{evG}) can be rewritten in a more explicit form. 
Using the definition (\ref{G}), formulae (\ref{Pi}) and (\ref{barPi}) we get 
two equivalent evolution equations for {\bf b} corresponding to the two
limits $x\to\pm\infty$:
\begin{mathletters}
\begin{equation}
\partial_t {\bf b} = -2i\omega(k){\bf b} + {\bf b}H_2\Upsilon(-
\infty,\infty;k)H_2 
+ H_1\Upsilon(-\infty,\infty;k)H_2,
\label{ev_b1}\end{equation}
\begin{equation}
\partial_t {\bf b} = -2i\omega(k){\bf b} - H_1\overline{\Upsilon}(-
\infty,\infty;k)H_1{\bf b} 
- H_1\overline{\Upsilon}(-\infty,\infty;k)H_2.
\label{ev_b2}\end{equation}
\end{mathletters}
Similarly, 
\begin{mathletters}
\begin{equation}
\partial_t \overline{{\bf b}} = 2i\omega(k)\overline{{\bf b}} 
+ \overline{{\bf b}}H_1\Upsilon(-\infty,\infty;k)H_1 + H_2\Upsilon(-
\infty,\infty;k)H_1,
\label{ev_barb1}\end{equation}
\begin{equation}
\partial_t \overline{{\bf b}} = 2i\omega(k) \overline{{\bf b}} 
- H_2\overline{\Upsilon}(-\infty,\infty;k)H_2 \overline{{\bf b}}  
- H_2\overline{\Upsilon}(-\infty,\infty;k)H_1.
\label{ev_barb2}\end{equation}
\end{mathletters}

Evolution of  zeros $k_j$ and $\overline{k}_j$ of $\det\Phi_+(k)$ and 
$\det\Phi^{-1}_-(k)$, 
respectively,  is derived  by differentiation of the determinants. 
In the integrable limit zeros do not depend on $t$. Then, for instance,
\[
\frac{\text{d} k_j}{\text{d} t} = -\left. \frac{\partial_t \det 
\Phi_+(k)}{\partial_k \det \Phi_+(k)}
\right|_{k=k_j} = -\left.\frac{ \text{Tr}\{\Pi(k)\}\det\Phi_+(k)}{\partial_k 
\det\Phi_+(k)}
\right|_{k=k_j}.
\]
To  calculate the r.h.s. in this formula recall that 
 the evolution functional $\Pi$ has simple pole at $k=k_j$ and that 
$\det\Phi_+(k)$ can be written as
\[
\det\Phi_+(k) = \det\phi_+(k)\det\Gamma(k) = \det\phi_+(k)\prod_{i=1}^s
\left(\frac{ k - k_i}{k-\overline{k}_i}\right)^{\nu_i},
\]
where $\det\phi_+(k)\ne0$. Simple calculations give 
\[
\frac{\text{d} k_j}{\text{d} t} = - \frac{\text{Tr} \{\text{Res} 
\Pi(k_j)\}}{\nu_j}.
\]
Here ``Res" denotes the residue of $\Pi(k)$ at $k=k_j$.  
Noticing that the l.h.s. does not depend on
$x$, we simplify the above equation setting $x\to\pm\infty$:
\begin{equation}
\frac{\text{d} k_j}{\text{d} t} = - \frac{\text{Tr}\{\text{Res} \Upsilon (-
\infty, \infty; k_j)H_2\}}{\nu_j}
\equiv \frac{\text{Tr}\{\text{Res} \Upsilon (-\infty, \infty; 
k_j)H_1\}}{\nu_j}.
\label{ev_k}\end{equation}
Similarly, using $\text{det}\Phi^{-1}_-$, we get
\begin{equation}
\frac{\text{d} \overline{k}_j}{\text{d} t} =  - 
\frac{\text{Tr}\{\text{Res}\overline{\Upsilon}
(-\infty,\infty;\overline{k}_j) H_2\}}{\nu_j}
\equiv \frac{\text{Tr}\{\text{Res} \overline{\Upsilon}(-
\infty,\infty;\overline{k}_j) H_1\}}{\nu_j}.
\label{ev_bark}\end{equation}

In derivation of the perturbation-induced evolution of the null vectors we 
will use the  following remarkable identities (written for the 
$x$-independent null vectors, see (\ref{def_P}))
\begin{equation}
\text{Res}\Pi(k_j) | P^{(j)}_l \rangle = - \frac{\delta k_j}{\delta t}| 
P^{(j)}_l \rangle, \quad
\langle \overline{ P}{}^{(j)}_l | \text{Res}\overline{\Pi}(\overline{k}_j)  
= \langle \overline{ P}{}^{(j)}_l | \frac{\delta \overline{k}_j}{\delta t},
\label{id_vec}\end{equation}
as well as other two identities: 
\begin{equation}
\Phi_+(k_j)e^{ik_j xA}\text{Res}\Pi(k_j) = 0,\quad
\text{Res}\overline{\Pi}(\overline{k}_j)
e^{-i\overline{k}_jxA}\Phi^{-1}_-(\overline{k}_j)= 0,
\label{id_phi}\end{equation}
which follow from (\ref{var_phi}). 
To verify the identities (\ref{id_vec}) one can proceed as follows. 
Introduce  functions $F_j^{(+)}(k) = (k - k_j)\Phi^{-1}_+(k)$ and 
$F_j^{(-)}(k) = (k -\overline{k}_j)\Phi_-(k)$. The regularization matrices 
$\Gamma^{-1}(k)$ and $\Gamma(k)$ have simple poles at $k=k_j$ and 
$k=\overline{k}_j$, respectively. Hence the introduced matrix functions are 
holomorphic in some neighborhoods of these points. Now compute the 
product 
\[
\left\{F^{(+)}_j(k) \frac{\delta \Phi_+(k)}{\delta t}| p^{(j)}_l 
\rangle\right\}_{k=k_j}  
= \left\{ (k-k_j)e^{ikxA}\Pi(k)e^{-ikxA}| p^{(j)}_l \rangle\right\}_{k=k_j}
\]
\[
= e^{ik_j xA}\text{Res}\Pi(k_j)e^{-ik_j xA}| p^{(j)}_l \rangle. 
\]
On the other hand, 
\[
\left\{F^{(+)}_j(k) \frac{\delta \Phi_+(k)}{\delta t}\right\}_{k=k_j}  
= \left\{ \frac{\delta }{\delta t}(k - k_j) I\right\}_{k=k_j}  
- \left\{\frac{\delta F^{(+)}_j(k)}{\delta t}\Phi_+(k)\right\}_{k=k_j} 
\]
\[
= -\frac{\delta k_j}{\delta t}I	 - \frac{\delta F^{(+)}_j(k_j)}{\delta 
t}\Phi_+(k_j).
\]
Multiplication by $| p^{(j)}_l \rangle $ of the latter formula and comparison 
with the former leads 
to the first identity in (\ref{id_vec}). The second one can be checked in 
similar way.  

Now let us derive evolution equations for the $x$-independent null vectors 
defined in (\ref{def_P}).  
To this goal we simply differentiate equation (\ref{vec_def})
\[
\frac{\delta }{\delta t} \left( \Phi_+(k_j)e^{ik_jxA}| P^{(j)}_l \rangle \right)
= \left\{\Phi_+(k)e^{ikxA} \Pi(k)|P^{(j)}_l\rangle\right\}_{k=k_j}
+ \frac{\delta k_j}{\delta t}\frac{\partial \Phi_+(k_j)}{\partial k}e^{ik_j 
xA}|P^{(j)}_l \rangle 
\]
\[
+\frac{\delta k_j}{\delta t}\Phi_+(k_j)ixAe^{ik_j xA} |P^{(j)}_l \rangle 
+\Phi_+(k_j)e^{ik_jxA}\frac{\delta |P^{(j)}_l \rangle  }{\delta t} = 0.
\]
Denote $\Pi_r(k_j)$ the value of the regular part of $\Pi(k)$ at $k =k_j$, 
\begin{equation}
\Pi_r(k_j) = \left\{\Pi(k) - \frac{\text{Res}\Pi(k_j)}{k - k_j}\right\}_{k=k_j}.
\label{Pi_r}\end{equation}
Then, using (\ref{id_vec}) and (\ref{id_phi})   to cancel out secular terms, 
we arrive at
\[
\Phi_+(k_j)e^{ik_j xA}\left\{ \frac{\delta |P^{(j)}_l \rangle  }{\delta t} + 
\Pi_r(k_j)
|P^{(j)}_l \rangle \right\} = 0.
\]
Using the same arguments as in section \ref{secRH} for derivation of 
the integrable  evolution we get, adding the latter,
\begin{equation}
\frac{\text{d} |P^{(j)}_l \rangle  }{\text{d} t} = - \Omega(k_j)|P^{(j)}_l 
\rangle 
- \Pi_r(k_j)|P^{(j)}_l \rangle.
\label{var_P}\end{equation}
Similarly,
\begin{equation}
\frac{\text{d} \langle \overline{P}{}^{(j)}_l |}{\text{d} t} = \langle 
\overline{P}{}^{(j)}_l | 
\Omega(\overline{k}_j)  + \langle \overline{P}{}^{(j)}_l | 
\overline{\Pi}_r(\overline{k}_j),
\label{var_barP}\end{equation}
where
\begin{equation}
\overline{\Pi}_r(\overline{k}_j) = \left\{\overline{\Pi}(k) - 
\frac{\text{Res}\overline{\Pi}(k_j)}
{k - \overline{k}_j}\right\}_{k=\overline{k}_j}.
\label{barPi_r}\end{equation}
In the case of the involution equation (\ref{var_barP}) 
is Hermitian conjugate to  (\ref{var_P}).  
Note that the l.h.s.'s of equations (\ref{var_P})  and (\ref{var_barP}) do not 
depend on $x$. 
Hence we can sent $x\to\pm\infty$ to considerably simplify these equations. 
Consider, 
for instance, (\ref{var_P}). Letting $x\to\infty$ and introducing the notations 
\[
|P^{(j)}_l \rangle = H_1 |P^{(j)}_l \rangle + H_2 |P^{(j)}_l \rangle \equiv
|P^{(j)}_l, 1\rangle + |P^{(j)}_l , 2\rangle
\]
and  $\Upsilon^{(1)} = H_1\Upsilon H_2$, $\Upsilon^{(2)} = H_2\Upsilon H_2$  we 
get the following system
\begin{mathletters}
\label{sys}
\begin{equation}
\frac{\text{d} |P^{(j)}_l, 1\rangle }{\text{d} t} = - i\omega(k_j) |P^{(j)}_l, 
1\rangle  
- \Upsilon^{(1)}_r(-\infty,\infty;k_j) |P^{(j)}_l , 2\rangle, 
\label{P_1}\end{equation}
\begin{equation}
\frac{\text{d} |P^{(j)}_l , 2\rangle }{\text{d} t} = \left\{
i\omega(k_j) - \Upsilon^{(2)}_r(-\infty,\infty;k_j)\right\} |P^{(j)}_l, 
2\rangle.
\label{P_2}\end{equation}
\end{mathletters}
Here $\Upsilon_r (k_j)$ is value of the regular part of the matrix
$\Upsilon(k)$  at 
$k=k_j$:
\begin{equation}
\Upsilon_r(k_j) = \left\{\Upsilon(k) - \frac{\text{Res}\Upsilon(k_j)}{k -
k_j}\right\}_{k=k_j}.
\end{equation}

When dealing with vector PDEs, i.e., for $n=N-1$, 
one can define  the polarization-phase parameters of  
vector solitons as quotients of components of the 
null vectors (note that the zeros are simple in this 
particular case). Let   $C^{(j)}_{l} = P^{(j)}_l/ P^{(j)}_N$, 
$l=1,...,n=N-1$, where $|P^{(j)}\rangle = (P^{(j)}_1, ... ,P^{(j)}_{N})^T$. 
Then  from (\ref{sys}) we obtain: 
\[
\frac{\text{d} C^{(j)}_l}{\text{d} t} = \left\{  -2i\omega(k_j) 
+ {\Upsilon_r}_{NN}(-\infty,\infty;k_j) \right\}C^{(j)}_l 
- {\Upsilon_r}_{lN}(-\infty,\infty;k_j),\quad l=1,...,n.
\]
This particular case ($n=N-1$) contains the vector NLS 
(\ref{vectorNLS}) and the complex modified KdV (\ref{CMKdV}) equations
as examples. In view of considerable importance of such vector
nonlinear PDEs, we formulate the result of this section in the following
theorem.
\bigskip

\noindent
{\bf Theorem:}
{\it  Let } 
\begin{equation}
iQ_t + V_0(Q,Q_x,Q_{xx},...) = \epsilon R(x,t,Q,Q_x,Q_{xx},...),
\label{evolution}\end{equation}
{\it be  a perturbed  nonlinear PDE associated with the 
$N\times N$ matrix Zakharov-Shabat spectral problem,}
\begin{equation}
\partial_x \Phi = ik[A,\Phi] + iQ(x,t)\Phi, \quad
A = \text{diag} (I_n,-1),
\label{spectral}\end{equation}
{\it where ($n=N-1$)}
\[
Q= \left( \begin{array}{cc}0 & {\bf q} \\ {\bf q}^\dagger & 0 
\end{array}\right),\quad
R = \left( \begin{array}{cc} 0 & {\bf r} \\ -{\bf r}^\dagger & 0   
\end{array}\right),\quad {\bf q} = \left(q_1, q_2, ..., q_n\right)^T, \quad 
{\bf r} = \left(r_1, r_2, ..., r_n\right)^T.
\]
{\it Here the matrix function} $V_0$ {\it represents the  limiting 
integrable evolution given by the dispersion relation} $\Omega(k) = 
i\omega(k)A$,
{\it while} $R$ {\it contains  the terms destroying integrability 
($\epsilon$ is a small parameter). Then  the perturbed evolution  
is equivalent to the following evolution of the spectral data:}
\begin{equation}
\frac{\text{d} k_j}{\text{d} t} =  - 
\epsilon\text{Res}\Upsilon_{NN}(t,k_j),
\label{equ1}\end{equation}
\begin{equation}
\frac{\text{d} C^{(j)}_l}{\text{d} t} =  \left\{ -2i\omega(k_j) + 
\epsilon{\Upsilon_r}_{NN}(t,k_j)\right\}
C^{(j)}_l - \epsilon{\Upsilon_r}_{lN}(t,k_j), \quad l=1,...,n,
\label{equ2}\end{equation}
\begin{equation}
\frac{\partial  b_l(k) }{\partial t}= 
\left\{ -2i\omega(k) + \epsilon\Upsilon_{NN}(t,k)\right\}
b_l(k) + \epsilon\Upsilon_{lN}(t,k), \quad l=1,...,n.
\label{equ3}\end{equation}
{\it Here}
\begin{equation}
\Upsilon(t,k) =  \int\limits_{-\infty}^\infty\text{d}\,x 
e^{-ikxA}\Phi_+^{-1}(x,t,k)
R(x,t)\Phi_+(x,t,k)e^{ikxA},
\label{ups}\end{equation}
\[
\Upsilon_r(t,k_j) =  \left\{\Upsilon(t,k) 
-\frac{\text{Res}\Upsilon(t,k_j)}{k - k_j}\right\}_{k=k_j}.
\] 
{\it The matrix function} $\Phi_+(x,t,k)$ {\it solves the spectral problem
(\ref{spectral}) and the RH problem}
\[
\Phi_+^{\dagger}(x,t,k)\Phi_+(x,t,k) = e^{ikxA} G(k,t) e^{-ikxA}, \quad
 k\in \text{Re}, 
\]
\[
\Phi_+(k)\to I,\quad k\to\infty
\]
{\it of analytic factorization of matrix} $G(k,t)$,
\[
G(k,t) = \left( \begin{array}{cc} I_n  & {\bf b}(k,t) \\ {\bf 
b}^\dagger(k,t)& 1  \end{array}\right),
\quad {\bf b} = \left(b_1,b_2,...,b_n\right)^T.
\]
{\it The Riemann-Hilbert problem has (simple) zeros}  $k_j(t)$, 
$j=1,...,s$, {\it given by } $a(k_j,t) = \det\Phi_+(x,t,k_j) = 0$. 
{\it The vector-columns} $ |C^{(j)} (t)\rangle = \left(C^{(j)}_1(t), 
C^{(j)}_2(t),...,C^{(j)}_n(t),1\right)$ {\it are the null vectors of the 
matrices} $\Phi_+(x,t,k_j)e^{ik_jxA}$:
\[
\Phi_+(x,t,k_j)e^{ik_jxA} |C^{(j)}(t)\rangle = 0,\quad j=1,...,s.
\]
{\it The initial spectral data are obtained  via  
solution of the spectral equation (\ref{spectral}) and represent the 
spectral  characterization of the potential} $Q(x,0)$. 
{\it For real} $k$, {\it the spectral densities of radiation} 
 $b_l(k,t)$ {\it and the function}  $a(k,t)$ {\it satisfy  the following 
identity }
\[
\overline{a}a =1-\sum_{l=1}^n \overline{b}_lb_l.
\]
{\it The  potential ${\bf q}(x,t)$  is  reconstructed by the formula} 
\[
q_l(x,t) = 2\lim_{k\to\infty}k(\Phi_+)_{l,N}(x,t,k),\quad l=1,...,n.
\]

Some comments are necessary on the use of equations 
(\ref{equ1})-(\ref{equ3}) for the spectral data. These equations are {\it 
exact}, i.e., they account for the perturbation exactly. As a 
consequence, for the non-integrable PDE (\ref{evolution}),  these equations 
are non-closed ODEs. Equations (\ref{equ1})-(\ref{equ3}) are non-closed  
because they contain explicitly the matrix $\Phi_+(x,t,k)$, 
 solution of the RH problem, obtaining which requires 
knowledge of the spectral data. Therefore, equations 
(\ref{equ1})-(\ref{equ3}) serve as the generating equations for the 
perturbation expansion: expanding the spectral data into the asymptotic  
power series in $\epsilon$, one obtains the  sequence of 
closed  approximate ODEs for the spectral data.  On this way, one does not 
need to solve the RH problem -- the computations are algebraic.
In the next section we consider a single multi-component soliton as an 
example.

\section{Multi-component soliton under perturbations}
\label{exmpl}

In this section  we apply the theorem for construction of the perturbation 
theory for a single multi-component soliton. This can be done without 
specifying the dispersion relation $\omega(k)$ determining the  
evolution of the spectral data in the unperturbed PDE. 
Hence our results apply to all nearly integrable vector PDEs associated 
with the Zakharov-Shabat spectral problem.  Below we derive equations 
describing the evolution of the
soliton parameters and give formulae for the first-order radiation. 
For a single vector soliton given by formula (\ref{one-sol}),  
i.e., $q_l = 2i\eta\theta_l e^{i(\xi/\eta )z }\text{sech}z$, 
where $ z=2\eta(x-\bar{x})$, the 
regularization matrix $\Gamma$  has the form 
\begin{equation}
\Gamma = I - \frac{i\eta}{(k-\overline{k}_1)\cosh{z}}
\left( \begin{array}{cc} e^{-z}|\theta\rangle \langle\theta | & 
e^{i(\xi/\eta)z} 
|\theta \rangle \\ \langle \theta |e^{-i(\xi/\eta)z} & 
e^{z} \end{array}\right).
\label{gamma1}\end{equation}
Here $k_1=\xi+i\eta$ and we have used the vector 
notations $|\theta\rangle = (\theta_1,...,\theta_n)^T$ and $\langle \theta| 
= (\overline{\theta}_1,...\overline{\theta}_n)$. 
To simplify some of the calculations 
introduce the following basis (in the $n$-dimensional subspace) 
\[
|\theta^{(1)}\rangle = |\theta\rangle,\quad |\theta^{(2)}\rangle,\quad
...,\quad |\theta^{(n)}\rangle; \quad \langle 
\theta^{(l)}|\theta^{(m)}\rangle = \delta_{lm}.
\]   
The explicit form of the vectors $|\theta^{(l)}\rangle$ for $l=2,...,n$ 
will not be needed at all. Also we will use the basis \[
|e_1\rangle=(1,0,...,0)^T,\quad |e_2\rangle=(0,1,0,...,0)^T,\quad 
...,\quad |e_n\rangle=(0,...,0,1)^T.
\]
With the help of the unitary transformation matrix $\Xi$, defined as
\begin{equation}
\Xi = \left(\begin{array}{cc} B & 0 \\ 0 & 1\end{array}\right),\quad
B = \sum_{l=1}^n|e_l\rangle\langle\theta^{(l)}|,
\label{Xi}\end{equation}
the regularization matrix $\Gamma$ can be considerably simplified:
\begin{equation}
\widetilde{\Gamma} = \Xi\Gamma\Xi^{-1} = I -
\frac{i\eta}{(k-\overline{k}_1)\cosh{z}}
\left( \begin{array}{cc} e^{-z}|e_1\rangle \langle e_1 | & 
e^{i(\xi/\eta)z} |e_1 \rangle \\ \langle e_1 
|e^{-i(\xi/\eta)z} & e^{z} \end{array}\right).
\label{gamma2}\end{equation}
Note the evident property
$\widetilde{\Gamma}^{-1} = \widetilde{\Gamma}^\dagger$.

The transformation matrix $\Xi$ depends on $t$  
due to the perturbation-induced time dependence of 
the $\theta$-parameters, but does not depend on $x$. This simple fact 
allows us to use the transformation with $\Xi$ inside the integrals defining 
$\Upsilon$ (\ref{ups}). Taking into account that $\Phi_+=\Gamma$ for the 
pure soliton solution of the unperturbed PDE, in the first order we get
\[
\widetilde{\Upsilon} = 
\Xi\Upsilon\Xi^{-1} = \int\limits_{-\infty}^\infty\text{d}x\, e^{-ikxA}
\widetilde{\Gamma}^{-1}\widetilde{R}\widetilde{\Gamma} e^{ikxA}.
\]
Here we have defined
\begin{equation}
\widetilde{R} = \Xi R\Xi^{-1} = \left(\begin{array}{cc}
0 & B|r\rangle \\ -\langle r|B^\dagger & 0 \end{array}\right),
\label{Rmod}\end{equation}
where, for convenience of the presentation below, we have  changed the 
notation for the perturbation: $|r\rangle={\bf r}=(r_1,...,r_n)^T$.

In fact, we will need only one diagonal element
\[
\Upsilon_{NN} = \left(\Xi^{-1} \widetilde{\Upsilon}\Xi\right)_{NN}
= \int\limits_{-\infty}^\infty\text{d}x\, 
\left(\widetilde{\Gamma}^\dagger\widetilde{R}\widetilde{\Gamma}\right)_{NN}
\]
and the following non-diagonal matrix elements
\[
\Upsilon_{lN} = \left(\Xi^{-1} \widetilde{\Upsilon}\Xi\right)_{lN}
= \sum_{m=1}^n B^{-1}_{lm}\widetilde{\Upsilon}_{mN} = 
\theta_l \widetilde{\Upsilon}_{1N} 
+ \sum_{m=2}^n B^{-1}_{lm}\widetilde{\Upsilon}_{mN}
=\theta_l \widetilde{\Upsilon}_{1N} + F_l,
\]
where $l=1,...,n$. The second term simplifies as follows
\[
F_l= \int\limits_{-\infty}^\infty\text{d}x\, e^{-2ikx}
\sum_{m=2}^n B^{-1}_{lm}\widetilde{R}_{mN}\widetilde{\Gamma}_{NN}
=\int\limits_{-\infty}^\infty\text{d}x\, e^{-2ikx}\widetilde{\Gamma}_{NN}
\sum_{m=2}^n \langle e_l|B^{-1}|e_m\rangle
\langle e_m|B|r\rangle
\]
\begin{equation}
=\int\limits_{-\infty}^\infty\text{d}x\, e^{-2ikx}
\widetilde{\Gamma}_{NN}\left(r_l - \theta_l\langle \theta|r\rangle\right).
\label{Fj}\end{equation}
After simple calculations we get:
\begin{equation}
\Upsilon_{NN} = \frac{i}{4}\int\limits_{-\infty}^\infty\text{d}z\,
\text{sech}^2z \left(\frac{e^{-z}}{k-k_1}+\frac{e^z}{k-\overline{k}_1}
\right)\left(r_0(z) +\overline{r}_0(-z)\right),
\label{gamNN}\end{equation}
\[
\Upsilon_{lN} = \frac{\theta_l}
{8\eta}\int\limits_{-\infty}^\infty\text{d}z\,
\text{sech}^2z\frac{\exp\{-2ikx+i(\xi/\eta)z\}}{(k-k_1)(k-\overline{k}_1)} 
\biggl\{-4\eta^2\overline{r}_0(-z)
\]
\begin{equation}
+\left[e^{-2z}(k-\overline{k}_1)+e^{2z}(k-k_1)\right]^2
r_0(z)\biggr\} + F_l.
\label{gamjN}\end{equation}
Here we have used the notation
\begin{equation}
r_0 = e^{-i(\xi/\eta)z}\widetilde{R}_{1N} 
= e^{-i(\xi/\eta)z}\langle\theta|r\rangle 
=e^{-i(\xi/\eta)z}\sum_{l=1}^n\overline{\theta}_l r_l.
\label{r0}\end{equation}

\subsection{Evolution of the  soliton parameters}
Let us first derive  evolution equations for the soliton
parameters  $\eta$, $\xi$, $\bar{x}$, and 
$\theta_l$, $l=1,...,n$.  
Using the identity $-2ik_1x +i(\xi/\eta)z = z +2\eta\bar{x}-2i\xi\bar{x}$ 
and the definition $C_l = \theta_l e^{2(\eta-i\xi)\bar{x}}$ 
from section~\ref{secRH}, we obtain:
\begin{equation}
\text{Res}\Upsilon_{NN}(k_1) = \frac{i}{4}\int\limits_{-\infty}^\infty
\text{d}z\,e^{-z}\text{sech}^2z\left(r_0(z) + \overline{r}_0(-z)\right),
\label{resups}\end{equation}
\begin{equation}
{\Upsilon_r}_{NN}(k_1) = \frac{1}{8\eta}\int\limits_{-\infty}^\infty
\text{d}z\,e^z\text{sech}^2z\left(r_0(z) + \overline{r}_0(-z)\right),
\label{upsrNN}\end{equation}
\begin{equation}
{\Upsilon_r}_{lN}(k_1) = 
C_l\left\{-2i\bar{x}\text{Res}\{\Upsilon_{NN}(k_1)\}
+{\Upsilon_r}_{NN}(k_1) + J_0\right\} + f_l,
\label{upsrjN}\end{equation}
where 
\[
J_0 = \frac{1}{4\eta}\int\limits_{-\infty}^\infty
\text{d}z\,\text{sech}^2z\left(\cosh{z} + ze^{-z}\right)
\left(r_0(z) - \overline{r}_0(-z)\right)
\]
and
\[
f_l =F_l(k_1) =\frac{e^{2(\eta-i\xi)\bar{x}}}
{4\eta}\int\limits_{-\infty}^\infty
\text{d}z\,\text{sech}z \left(e^{-i(\xi/\eta)z}r_l 
- \theta_l r_0\right).
\]

Let us first consider the more involved derivation of 
equations for $\bar{x}$ and $\theta_l$. 
From equations (\ref{equ2}), (\ref{upsrNN}), and 
(\ref{upsrjN})  we get:
\begin{equation}
\frac{\text{d}C_l}{\text{d}t} = \left(-2i\omega(k_1) 
+ 2\epsilon i\bar{x}\text{Res}\Upsilon_{NN}(k_1) - \epsilon 
J_0\right)C_l - \epsilon f_l,
\label{dCjdt}\end{equation}
from which it follows that
\[
\frac{\text{d}|C_l|^2}{\text{d}t} =\Bigl[4\text{Im}\{\omega(k_1)\}
-4\epsilon\bar{x}\text{Im}\{\text{Res}\Upsilon_{NN}(k_1)\} - 
2\epsilon\text{Re}\{J_0\}\Bigr]|C_l|^2 - 
2\epsilon\text{Re}\{f_l\overline{C}_l\}.
\]
Recalling that $\sum_{l=1}^n|\theta_l|^2=1$ we obtain:
\[
\frac{\text{d}\bar{x}}{\text{d}t} = \frac{e^{-4\eta\bar{x} }}{4\eta}
\sum_{l=1}^n \frac{\text{d}|C_l|^2}{\text{d}t} - \frac{\bar{x}}{\eta}
\frac{\text{d}\eta}{\text{d}t},
\]
\[
\frac{\text{d}\theta_l}{\text{d}t} = \theta_l
\left(C_l^{-1}\frac{\text{d}C_l}{\text{d}t} 
-2\frac{\text{d}(\eta\bar{x})}{\text{d}t}
+2i\frac{\text{d}(\xi\bar{x})}{\text{d}t}\right).
\]
The rest calculations are straightforward substitutions and using   
the identity $\sum_{l=1}^n f_l\overline{C}_l = 0$,
which follows from the definitions of $C_l$, $r_0$, and $f_l$. 
After simple calculations one gets a system of 
equations  for the soliton parameters:
\begin{equation}	
\frac{\text{d}\eta}{\text{d}t} = 
-\frac{\epsilon}{2}\int\limits_{-\infty}^\infty
\text{d}z\,\text{sech}z\text{Re}\{r_0\}, 
\label{foreta}\end{equation}
\begin{equation}
\frac{\text{d}\xi}{\text{d}t} = 
-\frac{\epsilon}{2}\int\limits_{-\infty}^\infty
\text{d}z\,\text{sech}z\,\text{tanh}z\text{Im}\{r_0\}, 
\label{detadt}\end{equation}
\begin{equation}
\frac{\text{d}\bar{x}}{\text{d}t} = \frac{\text{Im}\{\omega(k_1)\}}{\eta}
-\frac{\epsilon}{4\eta^2}\int\limits_{-\infty}^\infty
\text{d}z\, z\text{sech}z\text{Re}\{r_0\}, 
\label{forx}\end{equation}
\[
\frac{\text{d}\theta_l}{\text{d}t} = i\theta_l\left\{
\frac{2\text{Im}\{\overline{k}_1\omega(k_1)\}}{\eta}
-\frac{\epsilon}{2\eta^2}\int\limits_{-\infty}^\infty
\text{d}z\,\text{sech}z\Bigl[\xi z\text{Re}\{r_0\}
+\eta(1-z\text{tanh}z)\text{Im}\{r_0\}\Bigr]\right\}
\]
\begin{equation}
+\frac{\epsilon}{4\eta}
\int\limits_{-\infty}^\infty \text{d}z\,\text{sech}z
\left(\theta_l r_0 - e^{-i(\xi/\eta)z}r_l\right).
\label{fortheta}\end{equation}

These equations can be compared with the adiabatic equations derived by 
Karpman [21] for the single scalar soliton. 
First, it is convenient to introduce the average phase 
$\bar{\delta}$ of the soliton by setting 
\[
\bar{\delta} = \sum_{l=1}^n|\theta_l|^2\delta_l.
\]
The evolution equation for the average phase then follows from 
equation (\ref{fortheta}):
\begin{equation}
\frac{\text{d}\bar{\delta}}{\text{d}t} = 
\frac{2\text{Im}\{\overline{k}_1\omega(k_1)\}}{\eta}
-\frac{\epsilon}{2\eta^2}\int\limits_{-\infty}^\infty
\text{d}z\,\text{sech}z\Bigl[\xi z\text{Re}\{r_0\}
+\eta(1-z\text{tanh}z)\text{Im}\{r_0\}\Bigr].
\label{fordelta}\end{equation}
Remarkably, the slow evolution of the soliton amplitude $\eta$, phase 
gradient $\xi$, position $\bar{x}$ and average phase $\bar{\delta}$ is 
given by  equations similar to those derived for the single  scalar 
soliton. The only trace of the vector nature of the soliton in equations 
(\ref{foreta})-(\ref{forx}) and (\ref{fordelta}) is that the  ``scalar" 
perturbation $r_0$
obtains by averaging the original vector perturbation as follows 
\[
r_0 = e^{-i(\xi/\eta)z}\sum_{l=1}^n\overline{\theta}_lr_l.
\]

The equation for $\theta_l = s_le^{i\delta_l}$ can be cast in the form of 
two separate equations, one for the polarization parameters $s_l$ and the 
other for the phases $\delta_l$. We get:
\begin{equation}
\frac{\text{d}s_l}{\text{d}t} = \frac{\epsilon}{4\eta}
\int\limits_{-\infty}^\infty \text{d}z\,\text{sech}z
\text{Re}\left\{s_lr_0 
-e^{-i(\xi/\eta)z-i\delta_l}r_l\right\},
\label{forsj}\end{equation}
\[
\frac{\text{d}\delta_l}{\text{d}t} = 
\frac{2\text{Im}\{\overline{k}_1\omega(k_1)\}}{\eta}
-\frac{\epsilon}{2\eta^2}\int\limits_{-\infty}^\infty
\text{d}z\,\text{sech}z\Bigl[\xi z\text{Re}\{r_0\}
+\eta(1-z\text{tanh}z)\text{Im}\{r_0\}\Bigr]
\]
\begin{equation}
+\frac{\epsilon}{4\eta}
\int\limits_{-\infty}^\infty \text{d}z\,\text{sech}z
\text{Im}\left\{r_0 
- s_l^{-1}e^{-i(\xi/\eta)z-i\delta_l}r_l\right\}.
\label{fordeltaj}\end{equation}
Note that the equation for $\delta_l$ contains $s_l$ in the denominator
as a reflection of the fact that the phase is not defined for 
the components which are not excited.

\subsection{First-order radiation}
Now let us consider the evolution of the spectral densities $b_l(k)$ of 
radiation. Taking into account radiation in the first-order  approximation 
amounts to solving the linearized regular Riemann-Hilbert problem (or 
the jump problem),
\[
\phi_+(k) -\phi_-(k) = \Gamma(k) 
\left(\begin{array}{cc} 0 & e^{2ikx}|b(k)\rangle \\ \langle b(k)|e^{-2ikx} 
& 0\end{array}\right) \Gamma^{-1}(k),
\]
where we have used the notation $|b\rangle = {\bf b} = 
(b_1,...,b_n)^T$. Solution of the above jump problem is 
obtained by integration and using the  normalization 
condition $\phi_\pm(k)\to I$ as $k\to\infty$. We obtain
\[
\phi(k) = I +
\frac{1}{2i\pi}\int\limits_{-\infty}^\infty\frac{\text{d}\ell}{\ell-k}
\Gamma(\ell)\left(\begin{array}{cc} 0 & e^{2i\ell x}|b(\ell)\rangle \\ 
\langle b(\ell)|e^{-2i\ell x} & 0\end{array}\right) \Gamma^{-1}(\ell).
\]
 
The contribution from radiation to the solution follows from the formula 
(\ref{Q}),
\[
q^{(\text{rad})}_l = -2\lim_{k\to\infty}k\phi_{lN}(k) 
=\frac{1}{i\pi}\int\limits_{-\infty}^\infty\text{d}k
\left\{\Gamma(k) \left(\begin{array}{cc} 0 & e^{2ikx}|b(k)\rangle \\ 
\langle b(k)|e^{-2ikx} & 0\end{array}\right) \Gamma^{-1}(k)\right\}_{lN}.
\]
As in derivation of the equations for the soliton parameters it is 
convenient to use the transformation with  the matrix $\Xi$. 
Using this transformation  and formula (\ref{gamma2}) one can easily simplify 
the 
formula for radiation contribution. We get $q^{(\text{rad})}_l=
q_l^{(\parallel)} + q_l^{(\perp)}$, where the ``parallel" and 
``perpendicular" parts of radiation are defined as follows
\begin{mathletters}
\label{qrad}
\begin{equation}
q_l^{(\parallel)} =
\frac{\eta\theta_l e^{i(\xi/\eta)z}}{i\pi}
\int\limits_{-\infty}^\infty\frac{\text{d}\lambda}{\lambda^2+1}
\Bigl\{(\lambda+i\tanh{z})^2 e^{i\lambda 
z}\langle\theta|g(\lambda)\rangle 
+ \text{sech}^2z e^{-i\lambda z}\langle 
g(\lambda)|\theta\rangle\Bigr\},
\end{equation}
\begin{equation}
q_l^{(\perp)} =
\frac{\eta e^{i(\xi/\eta)z}}{i\pi}
\int\limits_{-\infty}^\infty\frac{\text{d}\lambda}{\lambda-i}e^{i\lambda z}
(\lambda+i\tanh{z})(g_l(\lambda)- \theta_l\langle\theta|g(\lambda)\rangle). 
\end{equation}
\end{mathletters}
Here, for convenience, we have introduced the modified spectral 
parameter by setting \mbox{$k= \eta\lambda + \xi$} and the 
modified spectral densities of radiation:
\begin{equation}
g_l(\lambda) = e^{2i(\eta\lambda+\xi)\bar{x}}
b_l(\eta \lambda +\xi).
\label{glambda}\end{equation}
The separation of radiation into parallel and perpendicular parts is 
due the following facts. While the perpendicular part $q_l^{(\perp)}$ 
satisfies the orthogonality property
\begin{equation}
\sum_{l=1}^n \overline{\theta}_l q_l^{(\perp)} =0,
\label{property1}\end{equation}
the parallel part of the radiation is given by the same formula as 
the radiation of the scalar soliton (multiplied by 
$\theta_l$), but for the averaged spectral 
density $\langle\theta| g(\lambda)\rangle$.  

Consider the evolution equation (\ref{equ3}) for the spectral densities of 
radiation. In the first-order approximation the term with $\Upsilon_{NN}$ 
can be neglected. When deriving evolution equations for the modified 
spectral densities in the first-order approximation one must take into 
account only the fast (or ``integrable") evolution of the soliton parameters 
involved in the definition of $g(\lambda)$. In particular,  
the modified spectral parameter $\lambda$ is $t$-independent. The only 
parameter which has the fast $t$-dependence in (\ref{glambda}) is $\bar{x}$. 
By differentiation of (\ref{glambda}) and using (\ref{equ3}) and (\ref{forx})
we obtain
\begin{equation}
\frac{\partial g_l(\lambda)}{\partial t} = 
i\Omega_r(\lambda)g_l(\lambda)
+\epsilon\theta_l \Upsilon^{(\parallel)}(\lambda) + 
\epsilon\Upsilon^{(\perp)}_l(\lambda),
\label{forgj}\end{equation}
where 
\begin{equation}
{\Omega_r}(\lambda) = 
2\Bigl[\lambda\text{Im}\{\omega(k_1)\} + \text{Re}\{\omega(k_1)\}- 
\omega(\eta\lambda+\xi)\Bigr].
\label{Omega}\end{equation}
The last two terms in equation (\ref{forgj}) 
come from $\Upsilon_{lN}$ (\ref{gamjN}) (the 
second one is the contribution of $F_l$ (\ref{Fj})). They read
\begin{equation}
\Upsilon^{(\parallel)}(\lambda) = \frac{1}{2\eta(\lambda^2+1)}
\int\limits_{-\infty}^\infty\text{d}z\, e^{-i\lambda z}\Bigl[(\lambda - 
i\tanh{z})^2r_0 - \text{sech}^2z\,\overline{r}_0\Bigr],
\label{gampar}\end{equation}
\begin{equation}
\Upsilon^{(\perp)}_l(\lambda) = \frac{1}{2\eta(\lambda +i)}
\int\limits_{-\infty}^\infty\text{d}z\, e^{-i\lambda z}(\lambda -i\tanh{z})
(e^{-i(\xi/\eta)z}r_l - \theta_l r_0).
\label{gamperp}\end{equation}
Due to the definition of $r_0$ (\ref{r0}), 
the perpendicular component satisfies the identity
\begin{equation}
\sum_{l=1}^n\overline{\theta}_l \Upsilon^{(\perp)}_l = 0.
\label{propert2}\end{equation}

Integrating the equation for $g_l(\lambda)$ with 
$g_l(\lambda,t=0)=0$ and using the result in (\ref{qrad})
we arrive at the first-order correction to initially pure soliton 
solution. In this case we obtain:
\begin{mathletters}
\label{qrad1}
\[
q_l^{(\parallel)} =\epsilon
\frac{\eta\theta_l e^{i(\xi/\eta)z}}{i\pi}
\int\limits_{-\infty}^\infty\frac{\text{d}\lambda}{\lambda^2+1}
\Bigl\{(\lambda+i\tanh{z})^2 e^{i\lambda 
z +i{\Omega_r}(\lambda)t}\gamma^{(\parallel)}(\lambda) 
\]
\begin{equation}
+ \text{sech}^2z e^{-i\lambda 
z-i{\Omega_r}(\lambda)t}\overline{\gamma}^{(\parallel)}(\lambda)\Bigr\},
\end{equation}
\begin{equation}
q_l^{(\perp)} =\epsilon
\frac{\eta e^{i(\xi/\eta)z}}{i\pi}
\int\limits_{-\infty}^\infty\frac{\text{d}\lambda}{\lambda-i}
(\lambda+i\tanh{z})e^{i\lambda 
z+i{\Omega_r}(\lambda)t}\gamma^{(\perp)}_l(\lambda),
\end{equation}
\end{mathletters}
where
\[
\gamma^{(\parallel)} = \frac{1}{2\eta(\lambda^2+1)}
\int\limits_{-\infty}^\infty\text{d}z\, e^{-i\lambda z}\Bigl[(\lambda - 
i\tanh{z})^2 \hat{r}^{(+)}_0  - 
\text{sech}^2z\,\hat{r}^{(-)}_0\Bigr],
\]
\[
\gamma^{(\perp)}_l = \frac{1}{2\eta(\lambda +i)}
\int\limits_{-\infty}^\infty\text{d}z\, e^{-i\lambda z}(\lambda -i\tanh{z})
(e^{-i(\xi/\eta)z}\hat{r}_l - \theta_l 
\hat{r}^{(+)}_0),
\]
with 
\[
\hat{r}^{(+)}_0(z,t,\lambda) = 
\int\limits_{0}^t\text{d}\tau\,e^{-i{\Omega_r}(\lambda)\tau}
r_0(z,\tau),\quad 
\hat{r}^{(-)}_0(z,t,\lambda) = 
\int\limits_{0}^t\text{d}\tau\,e^{-i{\Omega_r}(\lambda)\tau}
\overline{r}_0(z,\tau),
\]
\[
\hat{r}_l(z,t,\lambda) =  
\int\limits_{0}^t\text{d}\tau\,e^{-i{\Omega_r}(\lambda)\tau}
r_l(z,\tau).
\]

\section{Conclusions}

In construction of the perturbation theory our main idea is to use the 
Riemann-Hilbert problem associated with the integrable PDE for the 
nonlinear transformation of the {\it perturbed} PDE to the spectral space.  
The evolution equations for the spectral data follow from the  evolution 
functional,  an additional object one needs to introduce into the IST theory 
to account for perturbations. For a single vector soliton, the equations 
describing evolution of the soliton parameters  and  first-order radiation 
are given in explicit form. The method is not restricted to the first order 
only. For instance, the second-order equations can also be derived. The 
perturbation theory can be applied for description of dynamics of the 
spatial optical solitons, soliton pulses in the multispecies Bose-Einstein 
condensates, soliton propagation in optical fibre with the account of the 
arbitrary  polarization of light pulses, and for many other applications of 
the multi-component soliton equations.  

In this paper we have restricted  the consideration to the 
Zakharov-Shabat spectral problem. However, the approach of this paper was 
successfully applied to other spectral problems as well [41-45].
There, the evolution  functional  was derived and the 
evolution equations for the spectral data were 
obtained. The overall result of this  and the previous works on the 
perturbation theory based on the Riemann-Hilbert problem is that this 
approach {\it always works}. The explicit
form of the evolution functional was the  same for all considered spectral 
problems and, moreover, it undergoes only insignificant changes 
in the  transition from the Cauchy  problem to an initial-boundary 
value problem [45].

\acknowledgements
The author is indebted to Professor E. V. Doktorov for stimulating 
discussions during the course of this work and his critical reading of the 
manuscript.  This research was supported in part  by the NRF 
of South Africa. 
 
\appendix
\section{Some comments on multiplicity of zeros}
Here we explore in more detail the algebraic and geometric multiplicities of  
zeros of $\det\Phi_+(k)$ and $\det\Phi^{-1}_-(k)$. For instance, consider  
the determinant
\begin{equation}
\det \Phi_+ = {J_+}_1\wedge... \wedge {J_+}_n\wedge{J_-}_{n+1}\wedge ... 
\wedge{J_-}_N.
\label{A1}\end{equation}
None of the columns  ${J_\pm}_l(x,k)$ is equal to zero 
(otherwise, due to uniqueness of solution to equation (\ref{lax1}) we 
would have ${J_\pm}_l= 0$ for all $x$). Moreover, the minors in the 
wedge products of columns of $J_+$ and $J_-$ satisfy linear homogeneous 
equations, which follow from (\ref{lax1}), section \ref{secRH},
\[
\partial_x  {J_+}_1\wedge... \wedge {J_+}_n = \left\{ ik\left( 
\hat{A}^{(n)} - \frac{n(n+1)}{2}
\hat{I}^{(n)}\right) + i\hat{Q}^{(n)}\right\}  
{J_+}_1\wedge... \wedge {J_+}_n, 
\]
\[
\partial_x  {J_-}_{n+1}\wedge ... \wedge{J_-}_N= \left\{ ik\left( \hat{A}^{(N-
n)} - \frac{(N-n)(N-n+1)}{2} \hat{I}^{(N-n)}\right) + i\hat{Q}^{(N-n)}\right\} 
\]
\[
\times {J_-}_{n+1} \wedge ... \wedge{J_-}_N.
\]
Here $\hat{M}^{(j)}$  denotes a super-matrix, whose action on the wedge products 
of  vector-columns is defined by  the rule
\[
\hat{M}^{(j)}\Psi_1\wedge...\wedge\Psi_j = \sum_{p=1}^j \Psi_1\wedge...\wedge 
M\Psi_p\wedge
...\wedge\Psi_j.
\]
Therefore, $ \text{rank}{J_+}_1\wedge... \wedge {J_+}_n = \text{rank}\,e_1\wedge 
...\wedge e_n = n$,
and $\text{rank}{J_-}_{n+1} \wedge ... \wedge{J_-}_N = 
\text{rank}\,e_{n+1}\wedge ...\wedge e_{N-n} = N-n$. 
Hence, the  only possibility for $\det\Phi_+(k)=0$ is the linear dependence of 
the columns in 
${J_+}H_1$ and ${J_-}H_2$, e.g., at least one of the columns  of $J_-H_2$ is 
given as a linear combination of the columns of $J_+H_1$. 

Let $\nu_j$ and $d_j$ 
be the algebraic and geometric multiplicities of the $j$-th zero $k_j$, i.e., 
\[
\det\Phi_+(k)=(k-k_j)^{\nu_j}\psi(k), \quad \psi(k_j)\ne 0,\quad d_j = N- 
\text{rank}\Phi_+(k_j).
\]
Writing down the Taylor expansion of $\Phi_+(k)$ about $k=k_j$ we 
immediately conclude that
\begin{equation}
\text{rank}\Phi_+(k_j) \ge N- \nu_j.
\label{A2}\end{equation}
Therefore, in general, the algebraic multiplicity is greater  than the 
geometric one; trivially, they coincide 
for simple zeros. Representation (\ref{A1}) gives
\[
\text{rank}\Phi_+(k_j)\ge \text{max} (n,N-n).
\]
Hence the geometric multiplicity satisfies
\begin{equation}
d_j \le  N - \text{max}(n,N-n).
\label{A3}\end{equation}

We consider only zeros whose algebraic multiplicity is equal to the 
geometric one.  
In particular, if $n= N-1$, there can be  only one  vector in the null space 
of $\Phi_+$, i.e., $d_j=1$. 
Hence, in this case, zeros of $\det\Phi_+(k)$  must be simple to satisfy the 
equal multiplicity condition. 
This condition requires that the only case of a multiple zero, 
$\det\Phi_+(k_j)=0$, of order $\nu_j$ 
is that there are  precisely   $\nu_j$ columns of $J_-(k_j)H_2$ and 
$J_+(x,k)H_1$  which are linear combinations of the columns of $J_+(k_j)H_1$ 
and $J_-(k_j)H_2$, respectively. 

The equal multiplicity condition can be guaranteed by  the following 
constraint  (imposed for some $(x,t)$) \begin{equation}
\text{rank} \left(\Phi_+,\; \frac{\partial \Phi_+}{\partial k}\right) = N.
\label{A4}\end{equation}
Indeed, for algebraic multiplicity $\nu_j$, by the Taylor expansion,   
(\ref{A4}) gives $\text{rank}\Phi_+(k_j) = N - \nu_j$. 
Conversely, if  $\text{rank}\Phi_+(k_j) = N - \nu_j$, 
for algebraic multiplicity $\nu_j$,  then there are at least $\nu_j$ columns in 
$\partial \Phi_+(k_j)/\partial k $ independent from the columns of 
$\Phi_+(k_j)$. 
Hence (\ref{A4}) also holds. Similar results are valid for multiplicities of 
zeros $\overline{k}_j$, $j=1,...,s$, of $\det \Phi^{-1}_-(k)$.   

\section{Properties of the regularization matrix}
Here we derive the regularization matrix $\Gamma(k)$  and prove its 
properties (in the main, we follow Refs.~13  and 58). 
Dependence on the co-ordinates $x$ and $t$ is not important for this 
purpose and  omitted. Consider one pair of zeros, say, $k_s$ and 
$\overline{k}_s$ of $\det\Phi_+(k)$ and $\det\Phi^{-1}_-(k)$, respectively. 
Let the vectors $| p^{(s)}_l \rangle $ and $\langle \overline{ p}{}^{(s)}_l | $, 
$l=1,...,\nu_s$, satisfying  
\begin{equation}
\Phi_+(k_s)| p^{(s)}_l \rangle = 0,\quad \langle \overline{ p}{}^{(s)}_l 
|\Phi^{-1}_-(\overline{k}_s) = 0,
\label{B0}\end{equation}
span  the respective null spaces. Construct the following rational matrix 
functions
\[
\chi_s(k) = I - \frac{k_s - \overline{k}_s}{k - \overline{k}_s} P_s, \quad 
\overline{\chi}_s(k)  = I + \frac{k_s - \overline{k}_s}{k - k_s} P_s,
\]
where
\[
P_s = \sum_{l,m = 1}^{\nu_s} | p^{(s)}_l \rangle \left( M^{-1}\right)_{lm} 
\langle \overline{ p}{}^{(s)}_m |, \quad M_{lm}  = \langle \overline{ 
p}{}^{(s)}_l | p^{(s)}_m \rangle
\]
and $P_s$ is a projector: $P_s^2 = P_s$, $\text{rank}P_s = \nu_s$.  It is 
easy to verify that $\overline{\chi}_s(k) $ is inverse to $\chi_s(k)$: 
$\chi_s(k)\overline{\chi}_s(k) = I$. 

The determinant of $\chi_s(k)$ is easily computed in some appropriate  
basis, where  the projector is represented by a diagonal matrix with $\nu_s$ 
ones and $N-\nu_s$ 
zeros on the diagonal. We get
\[
\det \chi_s(k) = \left( \frac{k - k_s}{k - \overline{k}_s}\right)^{\nu_s}.
\]

Hence, with such rational matrices we can factor out the $s$-th pair of 
zeros. 
Indeed, consider the products $\Phi_+(k)\chi^{-1}_s(k)$ and $\chi_s(k)\Phi^{-
1}_-(k)$.  
These matrix functions are holomorphic  in the upper and lower half planes, 
respectively 
(the poles are removable due to the identities (\ref{B0})). 
On the other hand, the determinants are non-zero for $k=k_s$ and 
$k=\overline{k}_s$, 
respectively;  thus one pair of zeros is factored out. By introducing a 
sequence of such matrices,
\begin{equation}
\chi_j(k) = I - \frac{k_j- \overline{k}_j}{k - \overline{k}_j} P_j, \quad 
\overline{\chi}_j(k)  = I + \frac{k_j - \overline{k}_j}{k - k_j} P_j,\quad 
j=1,...,s,
\label{B1}\end{equation}
where  $\overline{\chi}_j(k) = \chi^{-1}_j(k)$, we factor out  all zeros 
using the regularization matrix $\Gamma$ and its inverse, where
\begin{equation}
\Gamma(k) = \chi_1(k)\chi_2(k)\cdot ... \cdot\chi_s(k).
\label{B2}\end{equation}
The projector $P_j$ is  given by the following formula
\[
P_j = \sum_{l,m=1}^s | e^{(j)}_l \rangle \left( M^{-1} \right)_{lm} \langle 
\overline{e}{}^{(j)}_m |,
\quad M_{lm} = \langle \overline{ e}{}^{(j)}_l | e^{(j)}_m \rangle. 
\]
Here the vectors $|e^{(j)}_l \rangle$ and $\langle \overline{e}^{(j)}_l | $ 
are related to the basis vectors  of the null spaces of 
$\Phi_+(k_j)$ and $\Phi^{-1}_-(\overline{k}_j)$ by triangular equations  
(if scanned starting from $s$ down to $1$):
\begin{equation}
| p^{(j)}_l \rangle = \chi^{-1}_s(k_j)\cdot\chi^{-1}_{s-1}(k_j)\cdot ... 
\cdot\chi^{-1}_{j+1}(k_j)
| e^{(j)}_l \rangle, 
\label{B3a}\end{equation}
\begin{equation}
\langle \overline{ e}{}^{(j)}_l |\chi_{j+1}(\overline{k}_j)\cdot 
\chi_{j+2}(\overline{k}_j)
\cdot... \cdot\chi_s(\overline{k}_j) = \langle \overline{ p}{}^{(j)}_l |.
\label{B3b}\end{equation}
Due to $P_j^2= P_j$, these vectors satisfy the identities:
\begin{equation}
\chi_j(k_j)| e^{(j)}_l \rangle = 0,\quad \langle \overline{e}{}^{(j)}_l 
|\chi^{-1}_j(k_j) = 0,\quad
l=1,...,\nu_j.
\label{B4}\end{equation}

The  regularization matrix $\Gamma(k)$ can be made parameterized entirely 
by the vectors from the null spaces. Indeed, let us decompose $\Gamma(k)$ 
and the inverse matrix into the partial fractions:
\begin{equation}
\Gamma(k) = I - \sum_{j=1}^s\frac{\overline{B}_j}{k - 
\overline{k}_j},\quad
\Gamma^{-1}(k) = I + \sum_{j=1}^s\frac{B_j}{k - k_j},
\label{B5}\end{equation}
where due to (\ref{B3a}) and (\ref{B3b}) we have
\begin{equation}
B_j = \sum_{l=1}^{\nu_j}| p^{(j)}_l \rangle \langle v^{(j)}_l |, \quad
\overline{B}_j = \sum_{l=1}^{\nu_j}| \overline{v}{}^{(j)}_l \rangle \langle 
\overline{p}{}^{(j)}_l |.
\label{B6}\end{equation}
 From  (\ref{B5})  and the identity $\Gamma\Gamma^{-1} = 
\Gamma^{-1}\Gamma = I$ it follows that \begin{equation}
\Gamma(k_j)| p^{(j)}_l \rangle = 0,\quad 
\langle \overline{ p}{}^{(j)}_l | \Gamma^{-1}(\overline{k}_j) = 0,\quad 
l=1,...,\nu_j,\quad j=1,...,s.
\label{B7}\end{equation}
These are the equations defining the unknown vectors $|\overline{v}{}^{(j)}_l 
\rangle $ and 
$\langle v^{(j)}_l |$. Indeed, rewriting (\ref{B7}) we have
\[
| p^{(j)}_l \rangle   = \sum_{i=1}^s  \frac{1}{k_j - 
\overline{k}_i}\sum_{m=1}^{\nu_i}
| \overline{v}{}^{(i)}_m \rangle \langle  \overline{ p}{}^{(i)}_m | 
p^{(j)}_l \rangle,\quad
\langle \overline{ p}{}^{(j)}_l | =  - \sum_{i=1}^s \frac{1}{\overline{k}_j 
- k_i}\sum_{m=1}^{\nu_i}\langle \overline{ p}{}^{(j)}_l | p^{(i)}_m 
\rangle\langle v^{(i)}_m |.
\]
Inversion of these formulae gives
\begin{equation}
| \overline{v}{}^{(j)}_l \rangle  = \sum_{i=1}^s\sum_{m=1}^{\nu_i}| p^{(i)}_m 
\rangle 
\left(D^{-1}\right)_{im,jl},\quad
\langle v^{(j)}_l | = \sum_{i=1}^s\sum_{m=1}^{\nu_i}\left(D^{-
1}\right)_{jl,im}\langle 
\overline{ p}{}^{(i)}_m |.
\label{B8}\end{equation}
Here the matrix $D$ is defined by
\[
D_{im,jl} = \frac{\langle \overline{p}_m^{(i)} |  
p_l{}^{(j)} \rangle }{k_j - \overline{k}_i}.
\]
Substitution of (\ref{B8})  into (\ref{B5}) produces the needed formulae 
(\ref{Gammas}).

\end{document}